\documentclass[11pt]{article} 
\usepackage{graphicx}
\usepackage{amsmath,amssymb}
\usepackage{cancel}
\usepackage{cite}
\usepackage{color}
\newcommand{\bea}{\begin{eqnarray}}
\newcommand{\eea}{\end{eqnarray}}
\newcommand{\nn}{\nonumber}

\newcommand{\eq}[1]{Eq.~(\ref{#1})}

\newcommand{\Msusy}{M_{\textrm{SUSY}}}

\newcommand{\ps}{p\hspace{-0.44em}/\hspace{0.06em}}

\begin{document} 
 
\pagestyle{plain}
\begin{flushright} 
TU-1002\\
CERN-PH-TH-2015-189
\end{flushright}
\vspace{5mm}
\begin{center}
{\LARGE{\bf Higgs-bosons couplings to quarks and leptons}}\\[5mm]
{\LARGE{\bf in the supersymmetric Standard Model}}\\[5mm]
{\LARGE{\bf with a gauge singlet}}\\[20mm]
{\large  Andreas Crivellin$^{1}$ 
and Youichi Yamada$^{2}$ }
\end{center}
\begin{center}
{\emph{
$^1$ Theory Division, CERN, CH-1211 Geneva 23, Switzerland \\[1.5ex]
$^2$Department of Physics, Tohoku University, Sendai 980-8578, 
Japan
}}
\end{center}\vskip 2cm

\begin{abstract}
The loop corrections to the couplings of Higgs bosons to quarks and charged leptons are calculated within supersymmetric versions of the Standard Model, extended by a gauge singlet. The effective couplings of the $SU(2)_L$ doublet and singlet Higgs bosons to quarks and leptons, induced by sfermion loops, are derived. Analytic expressions for the case of generic sfermion flavour mixing, including the complete resummation of all chirally-enhanced contributions are presented. These results are important in scenarios in which the mixing between singlet and doublet components of Higgs bosons is small, and the (pseudo) scalar component of the doublet is light. The calculated loop effects can have important consequences in flavour physics, especially for $\Delta F=2$ processes.
\end{abstract}

\newpage 
\setlength{\parskip}{1.02ex} 
%
\section{Introduction}
\label{sec:Intro}
%
The supersymmetric model with a gauge singlet scalar in addition to the usual 
 particle content of the Minimal Supersymmetric Standard Model~(MSSM) is a well motivated model, and was already proposed in the early
 days~\cite{Fayet,FZwirneretALL,Drees:1988fc} of studies of Supersymmetry~(SUSY), as a solution to the $\mu$ problem~\cite{KN}. 
It receives even more attention today in light of the recent Higgs boson discovery~\cite{EXPdiscov}.

These models were for sometime even the focus of Higgs-boson studies in SUSY before
 the Higgs boson was actually discovered with the relatively low value
 of mass of $\sim 126\,$GeV, which can be accounted for in the MSSM. 
The reason was the fact, noticed already quite early
 in Refs.~\cite{EspQuirW,MasMTPom,EspQuir,BasteroGilETALL}, that the singlet coupling to the Higgs doublets allows to break the link 
 of the Higgs quartic couplings to gauge couplings, typical of the MSSM.
Supersymmetric models with additional gauge singlets can therefore accommodate for a value of the SM-like Higgs mass much larger
 than that allowed in the MSSM, at least for small values of $\tan \beta$. 
Before the Higgs boson discovery, this fact reconciled SUSY aficionados 
 with the possibility that the LHC would find a Higgs boson far 
 heavier than that predicted by the MSSM (see for example Ref.~\cite{FranGori}).
It also possess the capability of reducing the amount of parameter fine
 tuning needed in order to obtain the correct Higgs mass 
(see for example~\cite{HPR,Ellwanger:2012ke,GhergvHMS}).  
Since then, it has become clear that a larger Higgs mass than that allowed 
 in the MSSM is also possible for moderate to large large values of 
 $\tan \beta$~\cite{Jeong2,Badziak:2013bda} due to the possible 
doublet-singlet mixing contributions to the physical Higgs masses.

In addition to the original scale-invariant model, called the 
Next-to-Minimal Supersymmetric Standard Model~(NMSSM), 
there exists other forms of the MSSM extended with a gauge singlet 
superfield. They differ by the singlet self interactions terms present in the
 superpotential. There is in particular the ``Minimal 
Non-minimal Supersymmetric Standard
 Model"~(MNSSM)~\cite{PanTam,PanPilaftsis} or nMSSM, and the
 PQ-NMSSM~\cite{KN,JSY,KNS,KChoi}. In all these cases, the only other particles to which the singlet superfield $S$
 couple are the Higgs doublet superfields $H_u$ and $H_d$. 
The models differ only in the way the Peccei--Quinn symmetry
 $U(1)_{\rm PQ}$~\cite{Fayet,PQ,JSY} of the superpotential is
 explicitly broken at the electroweak scale, i.e. either through a cubic
 term in $S$, or a linear one. The phenomenological studies of both models 
amounts to a sizable part of all beyond-SM 
analyses~\cite{REVIEWa,REVIEWb,BonaETAL,REVIEWc,REVIEWd,Oleg}.

Over the years, particular attention has been paid to the fact that the
 lightest pseudoscalar particle in these models, $a_1$, can be quite light 
(see for example Refs.~\cite{Dobrescu:2000yn,CDGW,Belyaev:2010ka,AM}, and 
references therein). 
Apart from small admixtures with the CP-odd neutral scalar components
 of $H_u$ and $H_d$, the mass eigenstate $a_1$ is mainly the pseudoscalar 
component of the singlet superfield $S$.  
The lightest CP-even neutral Higgs boson, $h_1$, may be equally light, 
depending on the values of various parameters of the model.  
Potentially enormous consequence arise from this fact for Higgs studies 
at the LHC~\cite{CDGW,Belyaev:2010ka,AM,explDMCOG4}.

The presence of light particles in the spectrum clearly affects also 
the physics of flavoured mesons, modifying therefore the MSSM searches
 at the high-luminosity frontier.
Indeed, the impact of very light Higgs bosons in $K$- and $B$-meson decays, 
has been the subject of intensive studies (see e.g. 
Refs.~\cite{BBMtoH,GHpseudosc,REVIEWb,BonaETAL,JMY}). 
Flavour physics will keep playing an important role in the exclusion or 
detection of such light states through searches at the Belle~II 
factory at SuperKEKB and with the LHCb program at CERN, in particular 
through the measurements of $B_d^0-\bar{B}_d^0$ and $B_s^0-\bar{B}_s^0$ 
mixing, and of the decay  $B_s \to \mu^+ \mu^-$. 
Direct searches for $a_1$ at the LEP \cite{LEP,ALEPH}, 
$B$ factories \cite{Domingo:2010am,Lees:2011wb,Lees:2012iw,Lees:2013vuj}, 
and the LHC \cite{Chatrchyan:2012am,Chatrchyan:2012cg} have 
already reduced  the parameter space of these models. 
Needless to say, direct searches of additional Higgs states will be able 
to probe whether singlet states are part of the Higgs sector or not.

It is therefore very important to know with a good precision 
 the couplings of the Higgs bosons to quarks and leptons 
 (of relevance for direct Higgs boson searches, flavour physics and 
also dark matter direct detection). 
Note that all Higgs mass eigenstates in this class of models, also those which are
 mainly singlet states, can couple to matter through their mixing with 
doublet Higgs states. 
Moreover, couplings of the singlet $S$ to matter fermion, even though 
they are vanishing at tree-level, are generated at the one-loop level 
as was first pointed out in 
Refs.~\cite{HodgPil,Hodgkinson:2008ei,BphysHodgPil}, and more recently 
in Ref.~\cite{JMY}. 
However, these effective singlet-fermion couplings have not attracted much attention. 
Although some existing codes for these models 
(such as {\tt NMSSMCALC}~\cite{NMSSMCALC} and {\tt SPheno}~\cite{SPheno}) 
include these couplings partially\footnote{In {\tt NMSSMTools}~\cite{CODESa} only the 
threshold corrections to the doublet Higgs Yukawa couplings but not the loop corrections 
to the singlet-quark couplings are implemented.}, 
no complete formula including resummation effects are implemented.
It is true that the experimentally found value of the Higgs mass 
 puts some emphasis on low values of $\tan \beta$, while these
 loop effects are maximized for the largest possible values of $\tan\beta$. 
Nevertheless, these effects cannot be neglected. 
They should be taken into account until the final embedding of the MSSM with 
 a singlet is --hopefully-- experimentally discovered. 
See, for example, Ref.~\cite{Ibanez} for the study of NMSSM with large $\tan\beta$. 

In this paper we plan to revisit the effective couplings of the singlet Higgs 
to leptons and quarks, induced by SUSY particle 
loops.\footnote{Note that the charged Higgs boson loops also give the $\tan\beta$ 
enhanced contribution in the NMSSM \cite{JMY}. We do not consider 
this contribution here, since it depends on the details of the Higgs potential.}
To this end we work in the approximation of large $\tan \beta$ where 
the loop effects are phenomenologically relevant.
This implies that a certain amount of fine tuning is needed to reproduce the measured value of the Higgs mass, like it is the case in the MSSM as well. With respect to the calculations in Refs.~\cite{HodgPil}, we drop the assumption of minimal flavour violation but rather include the effects of flavour-changing soft parameters. 
We also analytically re-sum all chirally-enhanced effects (as done in 
Ref.~\cite{Crivellin:2011jt} for the MSSM) for threshold corrections to Yukawa couplings 
and to the CKM matrix. While also Ref.~\cite{BphysHodgPil} worked in the MSSM with generic 
flavour structure\footnote{Ref.~\cite{BphysHodgPil} has presented numerical 
results only for the minimal flavour violation case. Note also that the calculation in 
Ref.~\cite{JMY} has included the $\tilde{t}_L-\tilde{c}_L$ mixing.},   
we include the wino and bino contributions which are not included 
 in Ref.~\cite{BphysHodgPil}. The results shown in this article are valid for all extension of the MSSM 
with a singlet, where the $\mu$ term is generated by the vacuum expectation 
value (VEV) of $S$, irrespectively of the singlet self-interaction terms allowed in the superpotential.\footnote{We assume a a CP-conserving Higgs potential. 
A recent study of the CP violation in the Higgs sector in NMSSM is 
seen in Refs.~\cite{JSLSenaha,MaggMuhll}.}

This article is organized as follows: In Sec.~2, we review the basic properties of SUSY standard models with an additional gauge singlet supermultiplet. In section~3, the self energies of quarks and leptons generated 
by SUSY loops are calculated. Section~4 deals with the effective couplings of the singlet Higgs which are expressed in terms of these self energy. Section~5 shows our numerical results and finally we conclude in Sec.~6.

\section{SUSY models with a gauge singlet}
\label{sec:NMSSM}
%
In this section we review the basics of models obtained adding a singlet field to 
 the MSSM particle content. In these models the superpotential is given by:
\begin{equation}
  W = W^{\rm MSSM}_{\rm Yuk} + \lambda\, S H_u \cdot H_d  + f(S) \,,  
\label{genSuperPW}
\end{equation}
where  $W^{\rm MSSM}_{\rm Yuk}$ contains the usual Yukawa structure:
\begin{equation}
W^{\rm MSSM}_{\rm Yuk} =  U^c  Y_U Q\cdot H_u - D^c Y_D Q\cdot H_d  - E^c Y_L L\cdot H_d \,,
\label{superPWYukawa}
\end{equation}
with $Q\equiv (U,D)^T$ and $L\equiv (N,E)^T$, the quark and lepton $SU(2)_L$ doublet superfields;
 $U^c$,  $D^c$,  and $E^c$, the $SU(2)_L$ singlet ones, and  $\cdot$ symbolizing the antisymmetric $SU(2)$ invariant product. The part of the superpotential denoted by $f(S)$ contains the singlet
 self interactions that identify the specific model in this class. 
Before proceeding, we give also the soft SUSY-breaking scalar terms needed to specify these models: 
\begin{equation}
  V_{\rm soft} =  V^{\rm MSSM}_{\rm soft-Yuk} + V^{\rm bilin}_{\rm soft} + \lambda A_\lambda  S H_u \cdot H_d + f_{\rm soft}(S)\,,
\label{eq:VSOFT}
\end{equation}
with
\begin{equation}
  V^{\rm trilin}_{\rm soft}  = 
 -\widetilde{U}^c {A}^{U}  \, \widetilde{Q} \cdot H_u
+\widetilde{D}^c {A}^{D}  \, \widetilde{Q} \cdot H_d
+\widetilde{E}^c {A}^{L}\, \widetilde{L} \cdot H_d 
+ {\rm H.c.} \, ,  
\label{eq:VSOFTtril}
\end{equation}
 where the $A$-term couplings $A^F_{ij}$ ($F=U,D,L$) are those often written as 
$(\pm)A^F_{ij}Y^F_{ij}$ in literature, and 
\begin{eqnarray}
  V^{\rm bilin}_{\rm soft} &\! =\! &
\widetilde{Q}^\ast \, \widetilde{m}^2_Q     \, \widetilde{Q} 
+\widetilde{U}^c    \, \widetilde{m}^2_{U^c} \, \widetilde{U}^{c\,\ast}
+\widetilde{D}^c    \, \widetilde{m}^2_{D^c} \, \widetilde{D}^{c\,\ast}
+\widetilde{L}^\ast \, \widetilde{m}^2_L \,     \widetilde{L} 
+\widetilde{E}^c    \, \widetilde{m}^2_{E^c} \, \widetilde{E}^{c\,\ast} 
\phantom{\frac{1}{2}}
\nonumber\\                        &   & 
+\widetilde{m}^2_{H_u} \, H_u^\ast H_u 
+\widetilde{m}^2_{H_d} \, H_d^\ast H_d 
+\widetilde{m}^2_{S}  \, S^\ast S\,. 
\label{eq:VSOFTmasses} 
\end{eqnarray}
The part $f_{\rm soft}(S)$ contains the mass and self couplings of the singlet scalar. Throughout this article, the same symbol is used for the superfields $H_u$, $H_d$, $S$ and their scalar component.
In all the above expression flavour indices have been suppressed. We shall return to this issue later in this section.

The gaugino mass terms:
\begin{equation}
-{\cal L}^{\rm gaugino}_{\rm soft} =
  \frac{1}{2}
 \left(M_3\, \widetilde{g}\widetilde{g}
      +M_2\, \widetilde{W}\widetilde{W} 
      +M_1\, \widetilde{B}\widetilde{B} \right)\,,
\label{eq:VSOFTgauginos}
\end{equation}
together with the terms in Eq.~(\ref{eq:VSOFT}) exhaust the list of soft SUSY-breaking terms in these models. 

As for parts of the potential describing the singlet self interaction, in the {\sf NMSSM},  $W(S)$ is usually chosen to be
\begin{equation}
 W(S) = \frac{1}{3} \kappa \, S^3\,, 
\label{eq:NsuperpWS}
\end{equation}
and $ V_{\rm soft}(S)$ is therefore 
\begin{equation}
  V_{\rm soft}(S) = 
 \frac{1}{3} \kappa A_\kappa S^3 
 + {\rm H.c.} \,.
\label{eq:NVsoftS} 
\end{equation}
A mass term $+m^\prime \zeta S^2$ is at times added 
to Eq.~(\ref{eq:NsuperpWS}), and a corresponding one 
$m^\prime B_\zeta S^2$ in Eq.~(\ref{eq:NVsoftS}) which break explicitly 
the scale invariance of the model.

In the {\sf nMSSM}, $W(S)$ contains a linear term in $S$:  
\begin{equation}
 W(S) = m^2 \xi S \ (+ m^\prime \zeta S^2) \,, 
\label{eq:PQsuperpWS1}
\end{equation}
where also an optional quadratic term in $S$, with a massive coupling $m^\prime$ can be added (see for example Ref.~\cite{KChoi}), and
\begin{equation}
  V_{\rm soft}(S) = m^2 \xi C_\xi S + ( m^\prime B_\zeta S^2) +{\rm H.c.} \,,
\label{eq:PQVsoftS2} 
\end{equation}
with $C_\xi$ and $B_\zeta$ being massive parameters.

No bilinear terms in $H_u \cdot H_d$ are present in the above superpotential, nor in the soft scalar potential terms. It is assumed in these models that such terms are 
generated once $S$ acquires a VEV $v_s \equiv \langle S\rangle$.
The effective $\mu$ and $B$ parameters are then given by
\begin{eqnarray}
 \mu_{\rm eff} &\equiv& {\lambda}\, v_s ,
\nonumber\\ 
 B_{\rm eff}   &\equiv& A_{\lambda} + \frac{1}{v_s}
\left. \frac{\partial f^*(S^*)}{\partial S^*}\right|_{S=v_s} \,, 
\label{muBeff}
\end{eqnarray}  
respectively. 

Note that the mass term $m^2$ in Eq.~(\ref{eq:PQsuperpWS1}) 
 may be generated spontaneously from non-renormalizable operators
 of the superpotential (in the Kim-Nilles mechanism~\cite{KN}) or of 
 the K\"ahler potential (in the Giudice-Masiero mechanism~\cite{GMasiero})
 involving the axion field, which acquires a VEV. 
There is in principle no particular constraint on this massive parameter $m$ 
to be linked with the electroweak scale.
A large tadpole term, however, tend to induce a very large value of 
 $v_S$. In such a case, an effective $\mu_{\rm eff}$, compatible with
 electroweak-symmetry breaking, can be obtained only for a tiny value of
 $\lambda$ (which implies a very small singlet-doublet mixing).

As is well known, the MSSM breaks explicitly the Peccei-Quinn~(PQ) symmetry
 through the $ \mu$ term. In the MSSM with a singlet it is possible to assign PQ charges to the 
 various fields in such a way to have $ W^{\rm MSSM}_{\rm Yuk} + \lambda\, S H_u \cdot H_d$ invariant under this
 symmetry~(see first column in Table~\ref{table:charges}).
{\small
\begin{table}[t]
\begin{center} 
\begin{tabular}{crr} 
\hline\hline                          &&\\[-1.9ex]       
                                      & 
$U(1)_{\rm PQ}\quad$                   & 
$U(1)_{\rm R}\quad$                    \\[1.01ex]\hline&&\\[-1.9ex]     
$S$                                   &
$-2\quad$                             &  
$2\quad$                              \\[1.01ex]\hline&&\\[-1.9ex]     
$H_u$                                 &
$1\quad$                              & 
$0\quad$                              \\[1.01ex]\hline&&\\[-1.9ex]     
$H_d$                                 &
$1\quad$                              &   
$0\quad$                              \\[1.01ex]\hline&&\\[-1.9ex]     
$(U^cQ)$                              &
$-1\quad$                             &  
$2\quad$                              \\[1.0001ex]\hline&&\\[-1.9ex]     
$(D^cQ)$                              & 
$-1\quad$                             & 
$2\quad$                              \\[1.0001ex]\hline&&\\[-1.9ex]     
$(E^cL)$                              & 
$-1\quad$                             & 
$2\quad$                              \\[1.0001ex]       
\hline\hline
\end{tabular} 
\caption{Peccei--Quinn and $R$-symmetry charges of the various fields.}
\label{table:charges}
\end{center}
\end{table} 
}
The explicit breaking of the PQ symmetry is thus shifted into $f(S)$.

It is clear that $ W^{\rm MSSM}_{\rm Yuk} + \lambda\, S H_u \cdot H_d$ also enjoys an $R$ symmetry with charges for the various fields listed in the second column of  Table~\ref{table:charges}.
We remind that the superpotential has $R$-charge $R(W) = 2$. Scalar, fermionic, and auxiliary component of the same chiral superfield $\Phi$ have different $R$-charges, with $R(\Phi_{\rm ferm})= R(\Phi_{\rm scal})-1$, and
 $R(\Phi_{\rm aux}) = R(\Phi_{\rm scal})-2$. Moreover, the fermionic component $\lambda_a$ of a gauge
 superfield $V_a$, has $R$-charge $R(\lambda_a) = 1$. Thus, the $R$ symmetry must be broken in the process of breaking SUSY, generating the $R$-violating mass terms for gauginos in Eq.~(\ref{eq:VSOFTgauginos}).

Through the VEV of $S$, induced by the electroweak-symmetry breaking, 
both  global $U(1)$ symmetries are spontaneously broken. In particular, 
the breaking of the PQ symmetry results in an 
 axion-like CP-odd scalar component of $S$, whose mass is
 proportional to $\kappa$~(or $\xi$ depending on the specific model) and therefore
 vanishing in the limit $\kappa\to 0$~(or $\xi\to 0$). The limit  $\kappa\to 
0$~(or $\xi\to 0$), in contrast, does not restore the $R$ symmetry, 
because there are other terms in the 
 Lagrangian that also break this symmetry.

In addition to the PQ and $R$ symmetric ones, there is also the MSSM limit, 
obtained for $\lambda, \kappa\to 0$ in the NMSSM (or  $\lambda, \xi\to 0$ 
in the nMSSM) while keeping $\mu_{\rm eff}$ finite. 
In this limit, $S$ tends to decouple from the model, giving rise to the superpotential and thus also to the scalar potential of the MSSM. 
However, the phenomenology may still be
 different from the one of the MSSM as the singlino (which is very weakly coupled
 for a finite but small $\lambda$) can be very light, i.e. the LSP. Therefore, expectations for DM and collider
 searches may differ substantially from those of the MSSM.

We close this overview of the models object of this paper with two subsections: 
one on the mixing of Higgs doublets and singlet, and 
one on the flavour basis we adopt in our investigation. 
%
\subsection{Mixing of Higgs bosons}
\label{HiggsMIX}

The singlet Higgs boson and the neutral components of the 
doublet Higgs bosons mix with each other. 
In order to study this mixing we expand the singlet field $S$ as
\begin{equation}
  S = v_s + \frac{1}{\sqrt{2}}\left\{ h_s + i a_s \right\} , 
\label{eq:Snormalization} 
\end{equation}
in analogy with $H_u^0$ and $H_u^0$ adopted in the MSSM 
\begin{eqnarray}
H_d^0 & = &  v_d + \frac{1}{\sqrt{2}}\left\{h_d + ia_d \right\}\,,
\nonumber \\
H_u^0 & = &  v_u + \frac{1}{\sqrt{2}}\left\{h_u + ia_u \right\}\,,
\label{HuHdNORMALIZ}
\end{eqnarray}
Here we have implicitly assumed that the global minimum of the Higgs scalar 
potential is realized by three real VEVs, $v_u$, $v_d$ and $v_s$.  

The presence of interaction terms among 
$H_u^0$, $H_d^0$ and $S$ in the scalar potential, such as
\begin{equation}
 V_{(H\!-\!S)} 
=\mu_{\rm eff}\lambda^*(S+S^*) ( v_d H_d^0 +v_u H_u^0 ) 
  -\lambda A_{\lambda}S ( v_u  H_d^0 + v_d  H_u^0 ) +\cdots  \,,
\label{eq:SHmixings}
\end{equation}
induces mass-mixing terms among these components. 
The mass terms in the basis (\ref{eq:Snormalization}, \ref{HuHdNORMALIZ}) 
are expressed as
\begin{equation}
 L^{\rm Higgs}_{\rm mass} = -\dfrac{1}{2}\left({\bf h}^T M_S^2 \,{\bf h}
+{\bf a}^T M_P^2 \,{\bf a}\right)\,,
\label{Higgsmasses}
\end{equation}
with
\begin{eqnarray}
{\bf h}&=\left( \begin{array}{c} h_d \\ h_u \\ h_s \end{array} \right)\,,
\qquad 
{\bf a}=\left( \begin{array}{c} a_d \\ a_u \\ a_s \end{array} \right)\,.
\label{Higgsvectors}
\end{eqnarray}
The explicit expressions for the  $3\times 3$ mass matrices $M_S^2$ 
and $M_P^2$ in the NMSSM can be found, for example, 
in Ref.~\cite{REVIEWc,REVIEWd}. 

The CP-even (CP-odd) mass eigenstates $h_1$, $h_2$ and $h_3$ 
($a_1$, $a_2$ and unphysical Nambu-Goldstone boson $G^0$) are given by 
\begin{equation}
O^S \left(  \begin{array}{c} h_2 \\ h_3 \\ h_1  \end{array} \right) =  
\left(  \begin{array}{c} h_d \\ h_u \\ h_s  \end{array} \right)\, ,\qquad
O^P \left(  \begin{array}{c} G^0 \\ a_2 \\ a_1  \end{array} \right)  = 
\left(  \begin{array}{c} a_d \\ a_u \\ a_s  \end{array} \right) \, ,
\label{Higgrotations}
\end{equation}
where $O^S$ and $O^P$ are 3$\times$3 orthogonal rotation matrices, such that
\begin{eqnarray}
(O^S)^T M_S^2\, O^S & = & {\rm diag}(m_{h_2}^2,m_{h_3}^2, m_{h_1}^2) \,,   
\nonumber \\
(O^P)^T M_P^2\,O^P & = & {\rm diag}(0, m_{a_2}^2 , m_{a_1}^2)\,.  
\label{eq:diagonaliz}
\end{eqnarray}
Note that we choose our conventions in such a way that if the lightest 
CP-even~(CP-odd) Higgs $h_1$~($a_1$) is mostly singlet-like we will have 
small angles in the mixing matrices.

In the CP-odd Higgs sector, it is often convenient to use the ``MSSM basis'' as 
an intermediate step, to separate the Nambu-Goldstone mode $G^0$ from 
physical states. The corresponding rotation is given by
\begin{equation}
\left( \begin{array}{c} a_d \\ a_u \\ a_s
 \end{array} \right) = O^P_\beta \left( \begin{array}{c} G^0 \\ A^0 \\ a_s
 \end{array} \right) = O^P_{\beta}O^P_{\theta_A} \left( \begin{array}{c} G^0 \\ a_2 \\ a_1
 \end{array} 
\right) \, , 
\label{MSSMbasis}
\end{equation}
 with 
\begin{equation}
O^P_{\beta} = \left(
 \begin{array}{ccc}
 \cos\beta & \sin\beta & 0 \\
-\sin\beta & \cos\beta & 0 \\ 
0 & 0 & 1  
 \end{array}\right) \, , \qquad 
O^P_{\theta_A} = \left(
 \begin{array}{ccc}
 1 & 0 & 0 \\
 0 & \cos\theta_A & \sin\theta_A \\
 0 & -\sin\theta_A & \cos\theta_A
 \end{array} \right)\, ,
\label{eq:AEigenstates}
\end{equation}
and $A^0$ is the mass eigenstate in the MSSM. 
With just a rewriting of the above equations, 
 the two pseudoscalars $a_1$ and $a_2$ are then given by
\begin{eqnarray}
 a_1 &=& \sin\theta_A A^0 + \cos \theta_A a_s \,,
\nonumber\\
 a_2 &=& \cos \theta_A A^0 -\sin\theta_A a_s \,.
\label{eq:massPSstates}
\end{eqnarray}
Note that our angle $\theta_A$ corresponds to the angle $\theta_A -\pi/2$ in the existing literature. 
In the MSSM limit, when $\lambda\to 0$ with fixed $\mu_{\rm eff}$, we have 
$\theta_A\to 0$. 
The angle $\theta_A$, however, can be very small also in other cases. For example, in the Peccei--Quinn limit, in which $a_1$ is light, it is given by (see for example~\cite{REVIEWd})
\begin{equation}
 \sin \theta_A = 
\frac{v \sin\beta \cos\beta}{\sqrt{v_s^2+ v^2 \sin^2 \beta \cos^2\beta}}\,, 
\label{eq:cosTHA} 
\end{equation}
which is greatly suppressed in the large $\tan\beta$ case. 

The rotation matrix for the CP-even Higgs states involve in general three different mixing angles. 
In the MSSM limit, it is 
\begin{equation}
 O^S = 
\left(
 \begin{array}{ccc}
  -\sin\alpha & \cos\alpha & 0 \\
  \cos\alpha & \sin\alpha & 0 \\
   0 & 0 & 1 
 
\end{array}\right) \, ,
\end{equation}
 with $h_2=h$ (the SM-like Higgs) and $h_3=H$ (the heavy CP-even Higgs), and the angle $\alpha$ defined as in the MSSM. In this same limit, if $H$, $A$, and $H^\pm$ are assumed to be heavy and nearly degenerate, the above matrix converges to:
\begin{equation}
O^S \to 
\left( \begin{array}{ccc} \cos\beta & \sin\beta & 0 \\
\sin\beta & -\cos\beta & 0 \\
0 & 0 & 1  \end{array}\right) \, .
\end{equation}

For reference, the physical state $H^\pm$ and the Goldstone mode
 $G^\pm$ of the charged Higgs bosons are the same as in the MSSM,
\begin{equation}
 \left(\begin{array}{c} G^\pm\\ H^\pm \end{array}\right)    = 
\left( \begin{array}{rr} \cos\!\beta & -\sin\!\beta \\ 
                  \sin\!\beta &  \cos\!\beta 
       \end{array} \right) 
 \left(\begin{array}{c} H_d^\pm\\ H_u^\pm \end{array}\right). 
\label{chargROT}
\end{equation}

\subsection{Flavour parameters/violations}
\label{FlavInLagr}
We work in the super-CKM basis, in which the tree-level mass matrices, i. e. the Yukawa couplings, of the quarks and leptons in the superpotential are diagonal~\cite{Hofer:2009xb,Crivellin:2009ar,
Crivellin:2011jt}. Then the Yukawa coupling matrices in the superpotential have the following forms:
\begin{equation}
(Y_U)_{ij} = Y^{u_i(0)} V^{(0)}_{ij}, \quad\quad 
(Y_D)_{ij} = Y^{d_i(0)}\delta_{ij},  \quad\quad 
(Y_L)_{ij} = Y^{\ell_i(0)}\delta_{ij} . 
\label{eq:superpWmatt}
\end{equation}
 where $V^{(0)}$ is the bare CKM matrix, arising from the misalignment
 between $Y_U$ and $Y_D$, which we have accommodated in the up-quark
 sector but one could have equally well shifted it into the down sector. 
The relation of the bare Yukawa couplings $Y^{f_i(0)}$ and $V^{(0)}$ to
 the physical fermion masses will be discussed in section 4.  
We also set $Y^{f_i(0)}$ ($f=u,d,\ell$) to be real and positive. 
The doublet superfield $Q$ is now defined as $Q = (V^{(0)\dagger} U, D)$. 
In this basis, the tree-level couplings of neutral Higgs bosons to
 quarks are, indeed, flavour diagonal. Similarly, in the trilinear SUSY-breaking terms in 
 Eq.~(\ref{eq:VSOFTtril}), $A^{U}$, $A^{D}$, $A^{L}$, have to be 
 understood as   
\begin{equation}
(A^U)_{ij}     = A^{u}_{ik}   V^{(0)}_{kj}\,,  \quad\quad 
(A^D)_{ij}     = A^{d}_{ij}           \, ,  \quad\quad 
(A^{L})_{ij} = A^{\ell}_{ij}         \, , 
\end{equation}
 where $A^{u}$, $A^{d}$, and $A^{\ell}$ are not necessarily diagonal and, 
in general, not even Hermitian. 

We now turn to the flavour violation of the sfermion (squarks and slepton) sector. 
The squark mass matrices in the super-CKM basis 
$(\widetilde{q}_{L1},\widetilde{q}_{L2},\widetilde{q}_{L3},\,
\widetilde{q}_{R1},\widetilde{q}_{R2},\widetilde{q}_{R3})$ 
are parametrized as 
\begin{equation}
\mathcal{M}^2_f\,=\,
\begin{pmatrix} 
m^2_{fLL} & m^2_{fLR} \\ 
(m^2_{fLR})^{\dagger} & m^2_{fRR} 
\end{pmatrix}
\label{eq:Deltas}
\end{equation}
with $m^2_{fXY}$ ($f=(u,d)$, $XY=(LL,RR,LR)$) being $3\times 3$ 
matrices in flavour space. 
The submatrices of Eq.~(\ref{eq:Deltas}) for down-type squarks are 
\begin{eqnarray}
m^2_{dLL} &=& \widetilde{m}^2_Q, \nonumber \\
m^2_{dRR} &=& \widetilde{m}^2_{D^c}, \nonumber \\
(m^2_{dLR})_{ij}&=&-v_d A^{d*}_{ji}\;-\;v_u\,\mu_{\rm eff}\, Y^{d_i(0)}\, 
\delta_{ij}\, .
\label{Deltad}
\end{eqnarray}
For up-type squarks, we have 
\begin{eqnarray}
m^2_{uLL} &=& V^{(0)}(\widetilde{m}^2_Q)V^{(0)\dagger}, \nonumber \\
m^2_{uRR} &=& \widetilde{m}^2_{U^c}, \nonumber \\
(m^2_{uLR})_{ij}&=&-v_u A^{u*}_{ji}\;-\;v_d\,\mu_{\rm eff}\, Y^{u_i{(0)}}\, 
\delta_{ij}\, .
\label{Deltau}
\end{eqnarray}
The mass matrix for the sleptons is obtained from $\mathcal{M}^2_d$ by 
replacing $(Q,D,A^{d})$ with $(L,E,A^{\ell})$, repspectively. We drop $O(v_{d,u}^2)$ 
terms in Eqs.~(\ref{Deltad}, \ref{Deltau}) 
in our approximation, as explained in Section 3. 
It should be noted that $m^2_{uLL}$ and $m^2_{dLL}$ 
are related by $SU(2)$ invariance.

\section{Quark and Lepton Self-Energies}
\label{sec:QLSelfEn}

In this sections we calculate the complete set of chirality changing one-loop quark and charged-lepton 
self energies induced by SUSY particles~(i.e. sfermions, gauginos and Higgsinos). 

\begin{figure}[t]
\begin{center}
\includegraphics[width=0.6\textwidth]{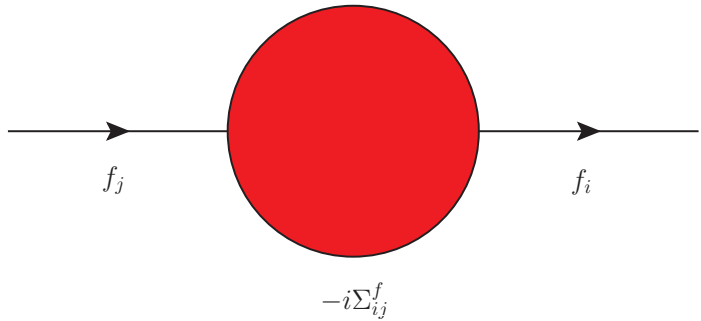}
\end{center}
\caption{The diagram shows our convention for the quark and
  lepton self energies $\Sigma^{f\,LR}_{ij}$. 
Here $i$ and $j$ are flavour indices and $f=(u,d,\ell)$.}
\label{fig:SE_def}
\end{figure}

We decompose the self-energies of quarks and leptons $-i\Sigma_{ij}^f (p)$ (see Fig.~\ref{fig:SE_def}) as 
\begin{equation}
 \Sigma_{ij}^f(p) = 
 \left[\Sigma_{ij}^{f\, LR}(p^2) +{\ps}\ \Sigma_{ij}^{f\, RR}(p^2)\right] P_R   
+ \left[\Sigma_{ij}^{f\,RL}(p^2) +{\ps}\ \Sigma_{ij}^{f\, LL}(p^2)\right] P_L 
\,.    
\label{Sigma:dec}
\end{equation}
with incoming (SM) fermion $f_j$ and outgoing fermion $f_i$. Here $f=(u,d,\ell)$ denotes the fermion type and $i$ and $j$ are flavour indices. 

Since we know that the SUSY particles are much heavier than the SM
 fermions, it is possible to expand $\Sigma_{ij}^f(p)$ in powers of
 $p/m_{\rm SUSY}$. For our purpose it is sufficient to evaluate the 
right-handed side of \eq{Sigma:dec} at $p^2=0$, i.e. at leading order in 
$p/m_{\rm SUSY}$. Furthermore, since we are only interested in
 chirally-enhanced effects related to Higgs-fermion couplings we only
 need the chirality-changing part of the self-energies:
\begin{equation}
 \Sigma_{ij}^{f\, LR}  \equiv \Sigma_{ij}^{f\, LR}(0)  = 
\Sigma_{ji}^{f\, RL\,\ast}(0) \,.
\end{equation}
We further assume that the masses of the SUSY particles in the loops
are sufficiently larger than the VEVs of doublet Higgs bosons,
i.e. $(v_d,v_u)\ll m_{\rm SUSY}$, and evaluate $\Sigma_{ij}^{f\, LR}$
to leading (first) order in $v/M_{\rm SUSY}$. 
We refer to this approximation as the decoupling limit since the remaining terms of
$\Sigma_{ij}^{f\, LR}$ do not vanish for 
$(M_{\rm SUSY},\mu_{\rm eff})\to\infty$. For
calculating $\Sigma^{f\,LR}_{ij}$ to leading order in $v/M_{\rm SUSY}$
the $SU(2)$-breaking elements of the SUSY mass matrices of the
sfermions, neutralinos and charginos (such as the left-right mixing of
sfermions or gaugino-higgsino mixing) are then not treated by the
mixing matrices but rather by mass insertions involving $v_d$ or
$v_u$. In this approach, all SUSY particles in the loops are the
$SU(2)_L$ gauge eigenstates. We also need to drop the $O(v_{d,u}^2)$ terms in the sfermion mass matrices, as is done in Eqs.~(\ref{Deltad}, \ref{Deltau}), 
in order to retain the non-decoupling terms only. 

The sfermion mass matrices still needs to be diagonalized due to possible 
flavour mixing in $m_{fLL}^2$ and $m_{fRR}^2$. 
Neglecting $m^2_{fLR}$ as explained above, the diagonalization is done 
as
\begin{eqnarray}
W^{f\dagger}\,\mathcal{M}^2_f\,W^f&=&
\textrm{diag}(m_{\tilde{f}_1^L}^2,m_{\tilde{f}_2^L}^2, 
m_{\tilde{f}_3^L}^2,m_{\tilde{f}^R_1}^2,m_{\tilde{f}^R_2}^2,m_{\tilde{f}^R_3}^2)\,,
\nonumber \\ 
W^f\,&=& \,\begin{pmatrix} W^{f\,L} & 0 \\ 0 & W^{f\,R} \end{pmatrix}\, .
\label{eq:Wmat}
\end{eqnarray}
The $3\times 3$ mixing matrices $W^{f\,L,R}$ take into account the flavour mixing originating from the terms $m^2_{fLL}$ and $m^2_{fRR}$, respectively. 
Note that the relations $m_{\tilde{u}_i^L}=m_{\tilde{d}_i^L}\equiv m_{\tilde{q}_i^L}(i=1,2,3)$ and 
$W^{u\,L} = V^{(0)} W^{d\,L}$ are fulfilled due to SU(2) invariance. 

For later convenience we introduce the abbreviations
\bea
\Lambda_{m\,ij}^{f\,LL} &=& (W^{f\,L})_{im}\,(W^{f\,L\star})_{jm}\, ,
\nonumber\\ 
\Lambda_{m\,ij}^{f\,RR} &=& (W^{f\,R})_{im}\,(W^{f\,R\star})_{jm}\, ,
\label{eq:Vmat}
\eea
where $i,j,m=1,2,3$. In Eq.~(\ref{eq:Vmat}) index $m$ is not summed over. 

\begin{figure}
\begin{center}
\includegraphics[width=0.5\textwidth]{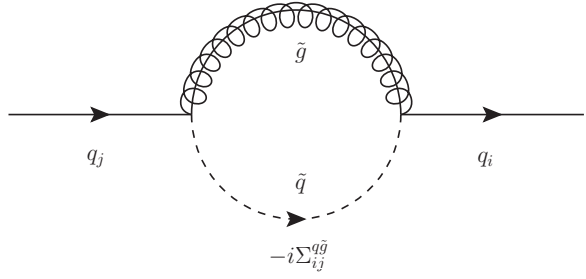}
\end{center}
\caption{Quark self-energy with gluino and squark as virtual particles.}
\label{fig:SelfEnergies1}
\end{figure}

\begin{figure}
\begin{center}
\includegraphics[width=0.5\textwidth]{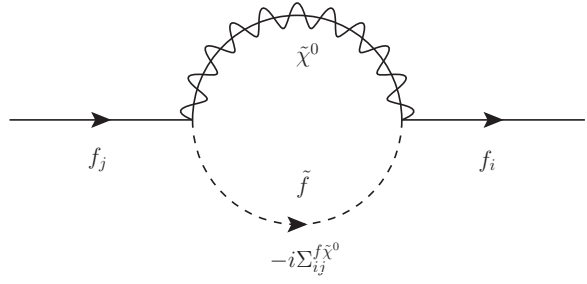}
\end{center}
\caption{Fermion (quark and lepton) self-energy with sfermions and
 neutralinos as virtual particles.}
\label{fig:SelfEnergies2}
\end{figure}
 
The self energies induced by sfermion loops resemble those in the
 MSSM by $\mu_{\rm eff}\leftrightarrow\mu$. Note that the elements $m^2_{fLR}$ in Eqs.~(\ref{Deltad}, \ref{Deltau}), 
inserted into sfermion propagators in the loops, may generate 
chirality-enhanced effects with respect to the tree-level masses 
if they involve the large VEV $v_u$~ ($\tan\beta$-enhancement for 
down-quark/lepton self-energies) or a trilinear 
$A^{f}$-term ($A^{f}_{ij}/(Y^f_{ij}\Msusy)$-enhancement). 

Below we list the relevant contribution (as calculated in Ref.~\cite{Crivellin:2011jt}).

We start from the gluino-squark contributions (see Fig.~\ref{fig:SelfEnergies1}). 
In our approximation, this contribution is proportional to $m^2_{qLR}$ and given by
\begin{eqnarray}
\Sigma_{fi}^{d \tilde{g}\,LR} &=& \dfrac{2\alpha_s}{3\pi}\, m_{\tilde
  g} \sum\limits_{j,k = 1}^3\; \sum\limits_{m,n =
  1}^3 \Lambda_{m\,fj}^{d\,LL}\; (m^2_{dLR})_{jk} \;
\Lambda_{n\,ki}^{d\,RR}\; C_0\! \left( m_{\tilde g}^2, m_{\tilde
  q_m^L}^2 ,m_{\tilde d_n^R }^2 \right) \,. 
\label{eq:gluinoSE}
\end{eqnarray}
Here $C_0$ is the standard three-point function~\cite{tHVelt,PassVelt} at vanishing momenta:
\begin{eqnarray}
 C_0(m_1^2, m_2^2, m_3^2) &\equiv  & 
 C_0(0,0, 0; m_1^2, m_2^2, m_3^2)
\nonumber \\
 & \equiv &
  \frac{-i}{\,\pi^2}  \int d^4 k
  \frac{1}{ \bigl[       k^2 - m_1^2       + i\epsilon \bigr]
            \bigl[ k^2 - m_2^2 + i\epsilon \bigr]
            \bigl[ k^2 - m_3^2   + i\epsilon \bigr]
          }           
\nonumber \\ 
   &\equiv&
   \frac{m_1^2 \,m_2^2 \,{\rm ln} ( m_1^2/ m_2^2 )  +
  m_2^2 \,m_3^2 \,{\rm ln} ( m_2^2/ m_3^2 ) +
  m_3^2 \,m_1^2 \,{\rm ln} ( m_3^2/ m_1^2 )
        }{(m_1^2 -m_2^2)(m_2^2 -m_3^2)(m_3^2 -m_1^2)}\,.
 \label{c0fnct}
 \end{eqnarray}

\begin{figure}
\begin{center}
\includegraphics[width=0.5\textwidth]{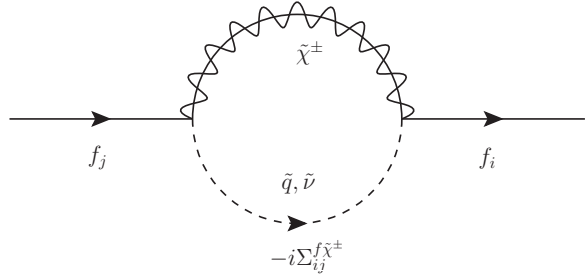}
\end{center}
\caption{Fermion self-energies with sfermions and charginos as virtual 
particles.}
\label{fig:SelfEnergies3}
\end{figure}

For the neutralino-sfermion contributions to lepton and quark 
self-energies shown in Fig.~\ref{fig:SelfEnergies2} we get
\begin{eqnarray}
\Sigma_{fi}^{\ell\tilde \chi^0 \,LR} &=& \dfrac{1}{16\pi^2}
\left\{\sum\limits_{j,k = 1}^3 \sum\limits_{m,n = 1}^3 g_1^2 M_1
\left(\Lambda_{m\,fj}^{\ell\,LL}\;\; (m^2_{\ell LR})_{jk} \;\;
\Lambda_{n\,ki}^{\ell\,RR}\right) C_0\!  \left( {\left| {M_1 }
  \right|^2,m_{\tilde \ell_{m}^L }^2 ,m_{\tilde \ell_{n}^R }^2 }
\right) \right.  \nonumber\\
&+& \sum\limits_{m = 1}^3 \left[ \dfrac{1}{\sqrt{2}g_2} M_W
  \sin\beta\;Y^{\ell_i(0)} \Lambda_{m\,fi}^{\ell \,LL} \left( g_2^2
  M_2 \mu_{\rm eff}\; C_0 \left( \left| M_2 \right|^2,\left| {\mu_{\rm eff} }
  \right|^2,m_{\tilde \ell_m^L }^2 \right) \right.\right.\nn\\
&-&\left. g_1^2 M_1 \mu_{\rm eff} C_0 \left( \left| M_1 \right|^2,\left| 
    \mu_{\rm eff}
        \right|^2,m_{\tilde \ell_m^L }^2 \right) \right) \nonumber\\
&+&\left.  \left.  g_1^2 \sqrt{2}\dfrac{M_W}{g_2} \sin\beta M_1 
\mu_{\rm eff}\;Y^{\ell_f(0) }
\Lambda_{m\,fi}^{\ell \,RR}\; C_0 \left( {\left| {M_1 }
  \right|^2,\left| {\mu_{\rm eff} } \right|^2,m_{\tilde \ell_m^R }^2 } \right)
\right] \right\}\,,
\end{eqnarray}
\begin{eqnarray}
\Sigma_{fi }^{d\tilde \chi^0 \,LR} &=& \dfrac{1}{16\pi^2}\left\{
\sum\limits_{j,k = 1}^3 {\sum\limits_{m,n = 1}^3 -\dfrac{1}{9}g_1^2
  M_1 \left({\Lambda_{m\,fj}^{d\,LL}\;\; (m^2_{dLR})_{jk} \;
    \Lambda_{n\,ki}^{d\,RR}}\right)} C_0\! \left( \left| {M_1}
\right|^2,m_{\tilde q_{m}^L }^2 ,m_{\tilde d_{n}^R }^2 \right) \right.
\nonumber\\
&+& \sum\limits_{m=1}^3 \left[ \dfrac{1}{\sqrt{2}g_2}M_W \sin\beta\;
  Y^{d_i(0)} \Lambda_{m\,fi}^{d\,LL}\left( g_2^2 M_2\; \mu_{\rm eff}\; C_0
  \left( \left| M_2 \right|^2,\left| \mu_{\rm eff} \right|^2,
   m_{\tilde q_m^L }^2
  \right) \right. \right.\nonumber\\
&+& \left. \dfrac{g_1^2 }{3} M_1 \mu_{\rm eff} C_0 \left( \left| 
M_1 \right|^2,\left|
  \mu_{\rm eff} \right|^2, m_{\tilde q_m^L}^2 \right) \right) \nonumber\\
&+& \left.  {\left.  \dfrac{1}{3}g_1^2 \sqrt{2}\dfrac{M_W}{g_2} 
     \sin\beta M_1 \mu_{\rm eff}\;
    Y^{d_f(0) } \Lambda_{m\,fi}^{d\,RR}\; C_0 \left( {\left| {M_1 }
      \right|^2,\left| {\mu_{\rm eff} } \right|^2,m_{\tilde d_m^R }^2 } 
  \right)
    \right]} \right\}\, .
\label{neutralino_SE}
\end{eqnarray}

Finally the chargino-sfermion contributions to lepton and down-quark 
self-energy (see Fig.~\ref{fig:SelfEnergies1}) are given by
\begin{eqnarray}
\Sigma_{fi}^{d\tilde \chi^\pm \,LR} &=& -\dfrac{Y^{d_i(0) }}{16\pi^2}
\mu_{\rm eff} \left[\delta_{i3}\,Y^{u_3(0)}\sum\limits_{m,n = 1}^3   
V^{(0)\star}_{3f}\;\Lambda_{m\,33}^{d\,LL}\; V^{(0)}_{33} \; 
(m^2_{uLR})^*_{33} \;   \Lambda_{n\,33}^{u\,RR} \; C_0\! 
\left( \left| \mu_{\rm eff} \right|^2 ,m_{\tilde q_m^L }^2,
m_{\tilde u_n^R }^2 \right) \right.
  \nonumber\\
&-& \left.  \sqrt 2 g_2 \sin \beta M_W M_2 \sum\limits_{m = 1}^3  
\Lambda_{m\,fi}^{q\,LL} C_0\! \left( {m_{\tilde q_m^L }^2 ,\left|
    \mu_{\rm eff} \right|^2 ,\left| {M_2 } \right|^2 } \right) \right]\,,
\nonumber \\
\Sigma_{fi }^{\ell\tilde \chi^ \pm \,LR} &=& \dfrac{\sqrt 2\, 
Y^{\ell_i(0) } }{16\pi^2} \mu_{\rm eff} g_2 \sin \beta M_W M_2 
\sum\limits_{m = 1}^3 \Lambda_{m\,fi}^{\ell \,LL} C_0\! 
\left( m_{\tilde \ell_m^L}^2 ,\left| \mu_{\rm eff} \right|^2 ,
\left| {M_2 } \right|^2 \right)\, .
\label{chargino-SE}
\end{eqnarray}
In Eqs.~(\ref{neutralino_SE}, \ref{chargino-SE}), $O(v_d)$ terms of the gaugino-higgsino mixing are neglected since they do not lead to chirally-enhanced contributions and cause unnecessary
complication due to their ultraviolet divergences. In addition, in the
higgsino-squark-squark contribution of Eq. (\ref{chargino-SE}) we have
further neglected small up-type Yukawa couplings of the first two
generations and multiple flavour-changes, instead of the full form
\begin{equation}
\Sigma_{fi}^{d\tilde{H}^\pm \,LR} = -\dfrac{Y^{d_i(0) }}{16\pi^2}\mu_{\rm eff}
\sum\limits_{f',j',j = 1}^3  V^{(0)*}_{f'f}\, Y^{u_{f'}(0)}  
\Lambda_{n\,f'j'}^{u\,RR} (m^2_{uLR})^*_{jj'} \; V^{(0)}_{ji'} \;
\Lambda_{m\,i'i}^{d\,LL}\;  C_0\! \left( \left| \mu_{\rm eff} \right|^2  ,
m_{\tilde q_m^L }^2 ,m_{\tilde u_n^R }^2 \right) \,.
\end{equation}
By using this approximation, we can find an analytic resummation formula taking into account all chirally-enhanced corrections~\cite{Crivellin:2011jt}. 

In contrast to down-type quarks, the up-type quark 
self-energies $\Sigma^{uLR}$ cannot be enhanced by $\tan\beta$. Nevertheless, an enhancement by 
$A^u_{ij}/Y^u_{ij}M_{\rm SUSY}$ is possible for the gluino and bino 
diagrams. These contributions are given as 

\begin{eqnarray}
\Sigma_{fi}^{u \tilde{g}\,LR} &=& \dfrac{2\alpha_s}{{3\pi }}\,
m_{\tilde g} \sum\limits_{j,k,j^\prime,f^\prime = 1}^3\;
\sum\limits_{m,n = 1}^3 V^{(0)}_{ff^\prime}\;\Lambda_{m\,f^\prime
  j^\prime}^{d\,LL} \;V^{(0)\star}_{jj^\prime}\;(m^2_{uLR})_{jk} 
\; \Lambda_{n\,ki}^{u\,RR}\; C_0\! \left( {m_{\tilde g}^2 ,
m_{\tilde q_m^L }^2 ,m_{\tilde u_n^R }^2 } \right) , 
\nonumber \\
\Sigma_{fi}^{u\tilde \chi^0 \,LR} &=& \dfrac{1}{16\pi^2}
\sum\limits_{m,n = 1}^3 \dfrac{2}{9}g_1^2 M_1 \;V^{(0)}_{ff^\prime}\; 
\Lambda_{m\,f^\prime j^\prime}^{d\,LL}
 V^{(0)\star}_{jj^\prime}\; (m^2_{uLR})_{jk} \;\Lambda_{n\,ki}^{u\,RR} 
C_0\! \left( {\left| {M_1 }
  \right|^2,m_{\tilde q_m^L }^2,m_{\tilde u_n^R }^2 } \right) \,.
\label{up-SE}
\end{eqnarray}

We then denote the sum of all contributions as
\begin{eqnarray}
\Sigma_{fi}^{d\,LR} &=& \Sigma_{fi}^{d\tilde{g}\,LR} +
\Sigma_{fi}^{d\tilde{\chi}^{0}\,LR} +
\Sigma_{fi}^{d\tilde{\chi}^{\pm}\,LR}\,,\nn\\
\Sigma_{fi}^{\ell\,LR} &=& \Sigma_{fi}^{\ell\tilde{\chi}^{0}\,LR} +
\Sigma_{fi}^{\ell\tilde{\chi}^{\pm}\,LR}\,, \nonumber\\
\Sigma_{fi}^{u\,LR} &=& \Sigma_{fi}^{u\tilde{g}\,LR} +
\Sigma_{fi}^{u\tilde{\chi}^{0}\,LR}\, .
\end{eqnarray}

As to be discussed later, the flavour off-diagonal pieces of 
$\Sigma^{f\,LR}_{fi}$ generate the flavour-changing couplings 
of the neutral Higgs bosons to quarks and leptons. 
In the case of the ``minimal flavour violation''  (MFV) 
with flavour-diagonal sfermion mass matrices, 
only charged higgsino contribution in Eq.~(\ref{chargino-SE}) 
cause flavour mixing originating from the CKM matrix. 

\section{Renormalization and threshold corrections} 

As already stated, in the fermion self-energies
(\ref{eq:gluinoSE}--\ref{chargino-SE}) the Yukawa couplings $Y^{f(0)}$
of fermions should be understood as the running ones of the
superpotential in the singlet-extended SUSY SM. These
couplings can be calculated from the physical masses of the fermions
by properly taking into account the SUSY threshold
 corrections. As is well known, the chirally-enhanced corrections to
the down-type quark masses may become numerically significant at large
$\tan\beta$~\cite{TBanks,Hempfling,HallRS,BFPT,CGNW,IsidoriR,
Hofer:2009xb} and must be resummed to all orders. 
In addition, the off-diagonal self-energies $\Sigma^f_{ij}$ cause rotation 
of the fermion mass eigenstates in the flavour space, 
which also generate difference between the bare CKM matrix $V^{(0)}$
of the superpotential and the physical one $V$.

In this section, we review the procedure of obtaining $Y^{f(0)}$ and 
$V^{(0)}$ from the SM running masses $m_{f_i}$ and CKM matrix $V_{ij}$, including 
the resummation of the chirally-enhanced corrections, following the 
results of Ref.~\cite{Crivellin:2011jt}. 

The running mass $m_{q_i}$ of the quark $q_i$ extracted from experiment 
using the SM prescription, is given by 
\begin{equation}
m_{q_i }  \;=\; v_q Y^{q_i(0)}  \,+\, \Sigma_{ii}^{q\,LR}\,,
\hspace{1cm} (q=u,d) \label{mq-Yq}\,.
\end{equation}
Here $\Sigma_{ii}^{q\,LR}$ is the flavour-diagonal piece of the
 self-energy calculated in the previous section.
Note that all terms in Eq.~(\ref{mq-Yq}) have to be evaluated at the
 same renormalization scale, i.e. the SUSY scale.

In the down-type quark sector, $\Sigma_{ii}^{d\,LR}$ is decomposed
 into the part which is proportional to a Yukawa coupling and the one
 which does not involve a Yukawa coupling, as
\begin{equation}
\Sigma_{ii}^{d\,LR} \;=\;
\Sigma_{ii\,\cancel{Y_i}}^{d\,LR} \, + \,\epsilon_i^{d}\,v_u\,\,Y^{d_i(0)}\,.
\label{eq:epsilon_b}
\end{equation}
This decomposition is possible if we restrict ourselves to the
decoupling limit where we have terms proportional to one power of
$Y^{d_i(0)}$ at most as can been see from~\eq{eq:gluinoSE},~(\ref{neutralino_SE}), and~(\ref{chargino-SE}). 
The second term of~\eq{eq:epsilon_b} gives chirally-enhanced
 corrections to the relation between the quark masses and the Yukawa couplings of the
superpotential, i.e. modifies this relation via a chirally-enhanced
threshold correction~\cite{TBanks,Hempfling,HallRS,BFPT,CGNW,IsidoriR,Hofer:2009xb}\footnote{For a 2-loop analysis of the threshold corrections to the relation between the Yukawa couplings and the quark masses see Ref.~\cite{2loop}.}. 
One automatically resums all chirally-enhanced corrections by inserting
\eq{eq:epsilon_b} into \eq{mq-Yq} and solving for $Y^{d_i(0)}$
\begin{equation}
Y^{d_i(0)} = \dfrac{m_{d_i} - \Sigma_{ii\,\cancel{Y_i}}^{d\,LR}}{v_d
  \left( {1 + \tan\beta \varepsilon_i^d } \right)}\,.
\label{md-Yd}
\end{equation}
The corresponding expressing for leptons follows trivially by 
replacing $d$ with $\ell$. In contrast, no such resummation is necessary for $Y^{u_i(0)}$ where 
only the contribution from $A$-terms can be significant:
\begin{equation}
Y^{u_i(0)} = \dfrac{m_{u_i} - \Sigma_{ii\,\cancel{Y_i}}^{u\,LR}}{v_u}\,.
\end{equation}

We now turn to the flavour-changing part of $\Sigma^{f\,LR}_{fi}$ which modifies the relation between the physical CKM matrix and the CKM matrix of the superpotential. In order to simplify the notation it is useful to define the quantity 
\begin{equation}
 \sigma^f_{ji}\,=\,\dfrac{\Sigma_{ji}^{f\,LR}}{\max\{m_{f_j},m_{f_i}\}}\, ,
\label{eq:sigdef}
\end{equation}
for $i\neq j$. The elements $\Sigma^{f\,LR}_{fi}$ contribute to the fermion mass
matrices and therefore require an additional rotation with respect to
the super-CKM basis to obtain the physical mass eigenstates of the
fermions ($\psi_i^{f\,L}$, $\psi_i^{f\,R}$)
\begin{equation}
\psi_i^{f\,L(R)}\to U^{f\,L(R)}_{ij}\psi_j^{f\,L(R)}\,. 
\end{equation}
To leading order in small ratios of the quark masses $m_{f_i}/m_{f_j}\ll 1$, 
$U^{f\,L}$ then reads~\cite{Crivellin:2008mq,Crivellin:2010gw,Crivellin:2010er}
\begin{equation}
\renewcommand{\arraystretch}{1.5}
U^{f\,L}  = \left( {\begin{array}{*{20}c}
1 & \sigma^f_{12} & \sigma^f_{13} \\
-\sigma^{f\star}_{12} & 1 & \sigma^f_{23} \\
-\left(\sigma^{f\star}_{13}\,-\,\sigma^{f\star}_{12}\,
\sigma^{f\star}_{23}\right) & 
-\sigma^{f\star}_{23} & 1 \\
\end{array}} \right)\,.
\label{DeltaU}
\end{equation}
The corresponding expressions for $U^{f\,R}$ are obtained from the
ones for $U^{f\,L}$ by the replacement $\sigma^f_{ji}\to\sigma^{f\star}_{ij}$.

Applying the rotations in~\eq{DeltaU} to the $\bar{u}_{iL} d_{jL}W^+$ vertex renormalizes the CKM matrix. 
The bare CKM matrix $V^{(0)}$ in Eq.~(\ref{eq:superpWmatt}) can be calculated in terms of the physical CKM matrix $V$ as
\begin{equation}
V^{(0)}  = U^{u\,L}\, V\, U^{d\,L\dag}\,.  
\label{CKM-0-ren}
\end{equation}
However, we have to take into account that $U^{d\,L}$ in Eq.~(\ref{CKM-0-ren}) depends on $V^{(0)}$ through the chargino loop contribution to $\Sigma^{d\,LR}$. In general, an iteration procedure
 is necessary to calculate $V^{(0)}$ using Eq.~(\ref{CKM-0-ren}). 
Nevertheless, in our approximation, we find a closed form of $V^{(0)}$. 
We first decompose $\sigma_{fi}^{d}$ as
\begin{equation}
\sigma_{fi}^{d} \;=\; \left\{\begin{array}{l}
\widehat{\sigma}^d_{f3}\,+\,\epsilon^d_{\textrm{FC}} V_{3f}^{\left(0
  \right)\star}V_{33}^{(0)}\,,\hspace{0.5cm}\textrm{i=3}\\
\widehat{\sigma}^d_{fi}\,,
\hspace{3.4cm} \textrm{i=1,2} \end{array}\right.\,,
\label{sigmahatd}
\end{equation}
so that $\widehat{\sigma}^d_{fi}$ does not depend on~(off-diagonal)
CKM elements and 
\begin{equation}
\varepsilon^d_{FC} = \dfrac{-1}{16\pi^2}\,\mu_{\rm eff}\,
\dfrac{Y^{d_3(0)}}{m_{d_3}} \sum\limits_{m,n = 1}^{3}
Y^{u_3(0)}\,\Lambda_{m\,33}^{d\,LL}\,(m^2_{uLR})^*_{33}\,
\Lambda_{n\,33}^{u\,RR}\, C_0 \left( \left| \mu_{\rm eff} \right|^2 ,
m_{\tilde q_m^L }^2 ,m_{\tilde u_{n}^R }^2 \right)\,.\label{eq:epsFC}
\end{equation}
In terms of the generalized Wolfenstein parametrization defined 
in the appendix of Ref.~\cite{Crivellin:2011jt}, we find
\begin{equation}
\renewcommand{\arraystretch}{1.5}
V_{}^{\left( 0 \right)}  = \left( {\begin{array}{*{20}c}
   {1 - \dfrac{{\left| {\widetilde{v}_{12} } \right|^2}}{2} + 
   i\,\widetilde{v}_{\rm{Im}} } & {\widetilde{v}_{12} } & 
   {\dfrac{{\widetilde{v}_{13} }}{{1 - \varepsilon_{FC}^d }}}  \\
   { - \widetilde{v}_{12}^{\star} } & {1 
 - \dfrac{{\left| {\widetilde{v}_{12} } \right|^2}}{2} 
   - i\,\widetilde{v}_{\rm{Im}} } & {\dfrac{{\widetilde{v}_{23} }}
 {{1 - \varepsilon_{FC}^d }}}  \\
 \dfrac{{\widetilde{v}_{12}^{\star} \widetilde{v}_{23}^{\star} 
 - \widetilde{v}_{13}^{\star} }}{{1 - \varepsilon_{FC}^{d\star} }} 
 & { - \dfrac{{\widetilde{v}_{23}^{\star} }}
 {{1 - \varepsilon_{FC}^{d\star} }}} & 1  \\
\end{array}} \right)\,.
\label{barCKM}
\end{equation}
with
\begin{eqnarray}
   \widetilde{v}_{12}  &=& v_{12}  \,+\, {\sigma}_{12}^{u} \,-\, 
                                \widehat{\sigma}^d_{12} \,,\hspace{1.5cm} 
   \widetilde{v}_{23}  \,=\, v_{23} \,+\, {\sigma}_{23}^{u} \,-\, 
                                \widehat{\sigma}^d_{23} \,,\nonumber\\[0.2cm]
 \widetilde{v}_{13}  &=& v_{13}  \,+\, {\sigma}_{13}^{u} 
 \,-\, \widehat{\sigma}_{13}^{d}
       \,+\, {\sigma}_{12}^{u} v_{23}  \,+\, 
 \left( \widehat{\sigma}_{12}^{d} \,-\, 
        {\sigma}_{12}^{u} \right)
      \widehat{\sigma}_{23}^{d} \,-\, v_{12} \widehat{\sigma}_{23}^{d}\,,
 \nonumber\\[0.2cm]
 \widetilde{v}_{\rm{Im}}  &=&  v_{12}\, 
{\mathop{\rm Im}\nolimits} \left[{\sigma}_{12}^{u}\, +\, 
 \widehat{\sigma}_{12}^{d}\right]\,-\,\textrm{Im}
 \left[{\sigma}^u_{12}\widehat{\sigma}^{d\star}_{12} 
 \right]\, ,
 \label{Vtilde2}
\end{eqnarray}
and
\begin{equation}
\renewcommand{\arraystretch}{1.5}
V_{}  = \left( {\begin{array}{*{20}c}
  {1 - \left| { {v}_{12} } \right|^2 /2 } & { {v}_{12} } &   { {v}_{13}  } \\
   { - {v}_{12}^{\star} } & {1 - \left| { {v}_{12} } \right|^2 / 2 } & 
{ {v}_{23}  }  \\
 {{ {v}_{12}^{\star} {v}_{23}^{\star}  - {v}_{13}^{\star} }}  & 
{ - {v}_{23}^{\star}  {}} & 1  \\
\end{array}} \right)\,.
\label{CKM_SM}
\end{equation}
We are now in a position to relate the self-energies $\Sigma_{ij}^{f\,LR}$ to corrections to the Higgs-fermion vertices and compute the effective Higges couplings as a function of SUSY breaking terms.

\section{Effective Higgs couplings to quarks and leptons}
\label{sec:higgs_vertex}

In this section we compute the effective Higgs-fermion couplings in
the MSSM with an additional gauge singlet superfield. For this purpose
 we first determine the couplings to the Higgs doublets ($H_u$ and
 $H_d$) and the singlet $S$, and go afterwards to the physical basis
 with diagonal Higgs mass matrices.

\subsection{Couplings to doublet Higgs bosons}

\begin{figure}
\begin{center}
\includegraphics[width=0.99\textwidth]{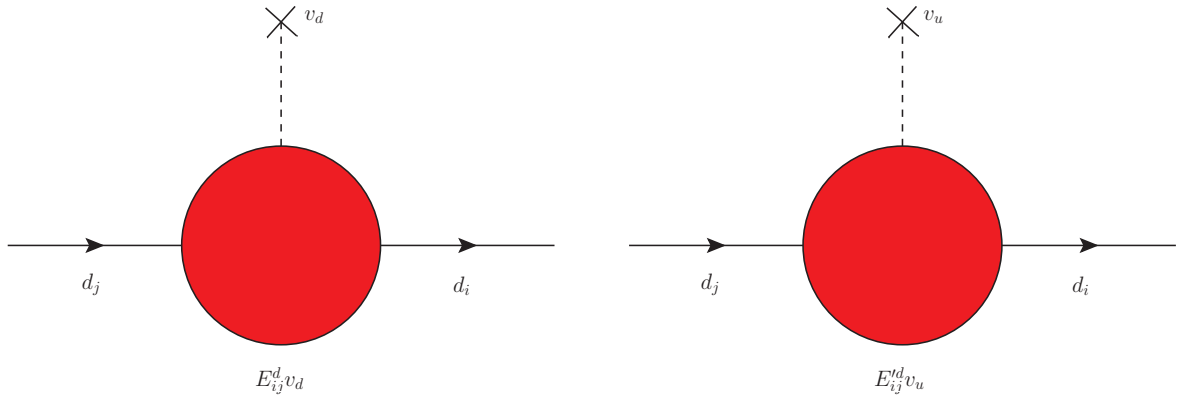}
\end{center}
\caption{Decomposition of the down-type quark self-energy into its 
holomorphic part $E^{d}_{ij}$ and its non-holomorphic part 
$E^{\prime d}_{ij}$. This decomposition is possible in the decoupling limit, i.e. if $\Sigma_{ij}^{f\,LR}$ is
 evaluated at leading order in $v/m_{\rm SUSY}$.
For charged leptons one simply has to replace $d$ with~$\ell$. }
\label{fig:SE_dec}
\end{figure}

The calculation of the effective couplings of the Higgs doublets $H_u$
and $H_d$ is the same as in the MSSM. We first decompose $\Sigma_{ij}^{f\,LR}$ as 
\begin{eqnarray}
\Sigma_{ij}^{(d,\ell)\,LR} &=& E_{ij}^{(d,\ell)} v_d 
+ E_{ij}^{\prime (d,\ell)} v_u,\, ,
\label{HoloDeco} \\
\Sigma_{ij}^{u\,LR} &=& E_{ij}^{u} v_u + E_{ij}^{\prime u} v_d,\, . 
\label{HoloDeco-up}
\end{eqnarray}
The first terms on the right-handed side of Eq.~(\ref{HoloDeco}) is
the holomorphic part, generated by the loop correction to coupling
$\bar{f}_{iL}f_{jR}H_d^*$ which exists already at the tree-level and
is only induced by $A$-terms in our approximation. 
In contrast, the second non-holomorphic term of Eq.~(\ref{HoloDeco}) comes from the
loop-generated effective $\bar{f}_{iL}f_{jR}H_u$ coupling involving
the effective $\mu_{\rm eff}$ term as shown in
Fig.~\ref{fig:SE_dec}. The term in Eq.~(\ref{HoloDeco}) proportional
to $v_u$ is always accompanied by a factor $v_s$ due to the PQ symmetry.
For Eq.~(\ref{HoloDeco-up}), similar discussion holds by exchanging 
$(H_d,v_d)$ and $(H_u,v_u)$. Note that $E_{ij}$ and $E^\prime_{ij}$ are general functions of $v_s$. 

The loop contributions to the effective couplings to doublet Higgs are then 
obtained by replacing $v_d$ and $v_u$ in Eq.~(\ref{HoloDeco}) 
(Eq.~(\ref{HoloDeco-up})) by $H_d^*$ ($H_d$) and $H_u$ ($H_u^*$), 
respectively. 

\subsection{Effective singlet-fermion couplings}

We are now in a position to calculate the loop-induced couplings of
the singlet Higgs $S$ to quarks and charged leptons resumming all
chirally-enhanced corrections. Note that also these loop-corrections
are directly related to quark and lepton self-energy generated once
the singlet acquires its VEV $v_s$ giving rise to $\mu_{\rm eff}$. 
Therefore, the effective couplings in our approach can also be
calculated in terms of the chirally-enhanced quark self-energies as
done before for the doublet Higgs couplings. For the couplings to the singlet we only consider down-type quarks and charged leptons since the contributions to up-type quark couplings 
are not chirally enhanced. We first derive the couplings in the super-CKM basis 
and relate them to the effective coupling in the physical basis with 
diagonal quark and lepton mass matrices. The effective Lagrangian 
governing the interactions of quarks and leptons with the singlet is given by 
\begin{equation}
L_{\rm eff} = \Gamma_{q_fq_i}^{h_s\,LR} h_s \bar{q}_{fL} q_{iR} 
+ i \Gamma_{q_fq_i}^{ a_s\,LR} a_s \bar{q}_{fL} q_{iR}  + {\rm h. c.} 
+ (q\to\ell) \, .
\end{equation}

\begin{figure}
\centering
\includegraphics[width=0.5\textwidth]{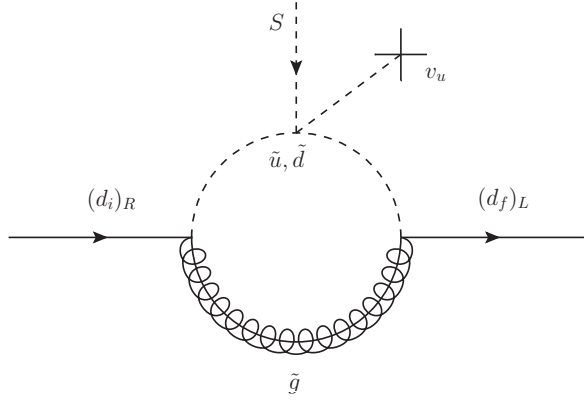}
\caption{Feynman diagram showing the gluino contribution to the effective 
singlet-down-quark couplings in the decoupling limit (i.e. in leading 
order of $v/m_{\rm SUSY}$).}
\label{fig:gluino}
\end{figure}

\subsubsection{Gluino and bino induced couplings}

For the effective quark-singlet coupling induced by a gluino shown in
 Fig.~\ref{fig:gluino} it is sufficient to replace $\mu_{\rm eff}$ in
 Eq.~(\ref{eq:gluinoSE}) by $\lambda$ in order to get the corresponding
 effective coupling to the singlet. The same is true for the pure bino contribution to the quark or lepton
 self-energy. The expression for the effective gluino induced coupling then reads:
\begin{equation}
\widehat\Gamma _{d_f d_i }^{\tilde g h_s\,LR}  = 
\widehat\Gamma _{d_f d_i }^{\tilde g a_s \,LR}
= -\dfrac{1}{\sqrt{2}} \dfrac{\lambda v_u}{\mu _{\rm eff} }  
E_{fi}^{\prime d\tilde g} \,. 
\end{equation}
Here the hat refers to the fact that the couplings are given in the
 super-CKM basis. The expressions for the bino contribution to effective quark or lepton
 couplings is simply obtained by exchanging the corresponding sub- and
 super-scripts.

\subsubsection{Chargino and neutralino induced couplings}

\begin{figure}
\centering
\includegraphics[width=0.48\textwidth]{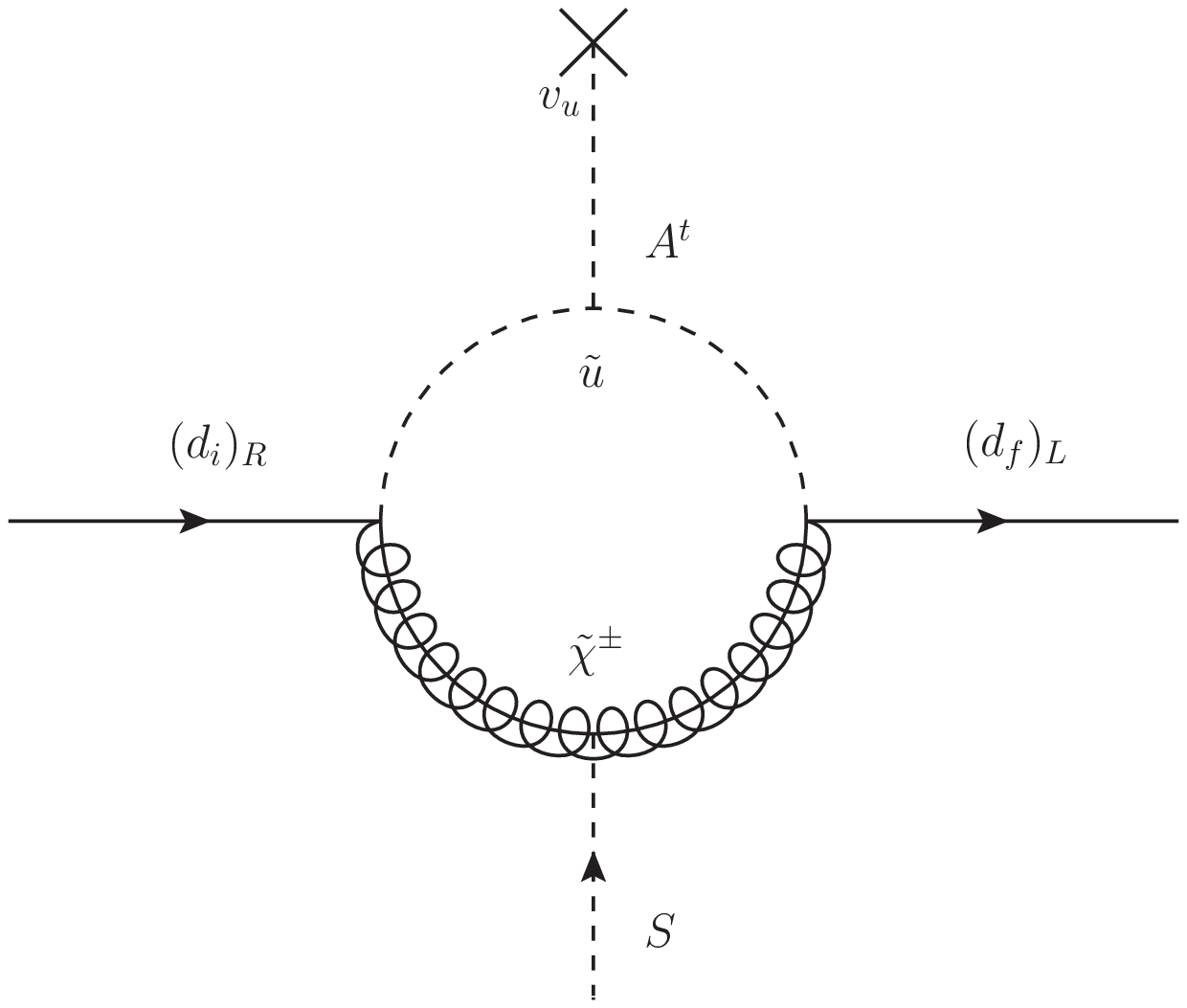}~~~
\includegraphics[width=0.48\textwidth]{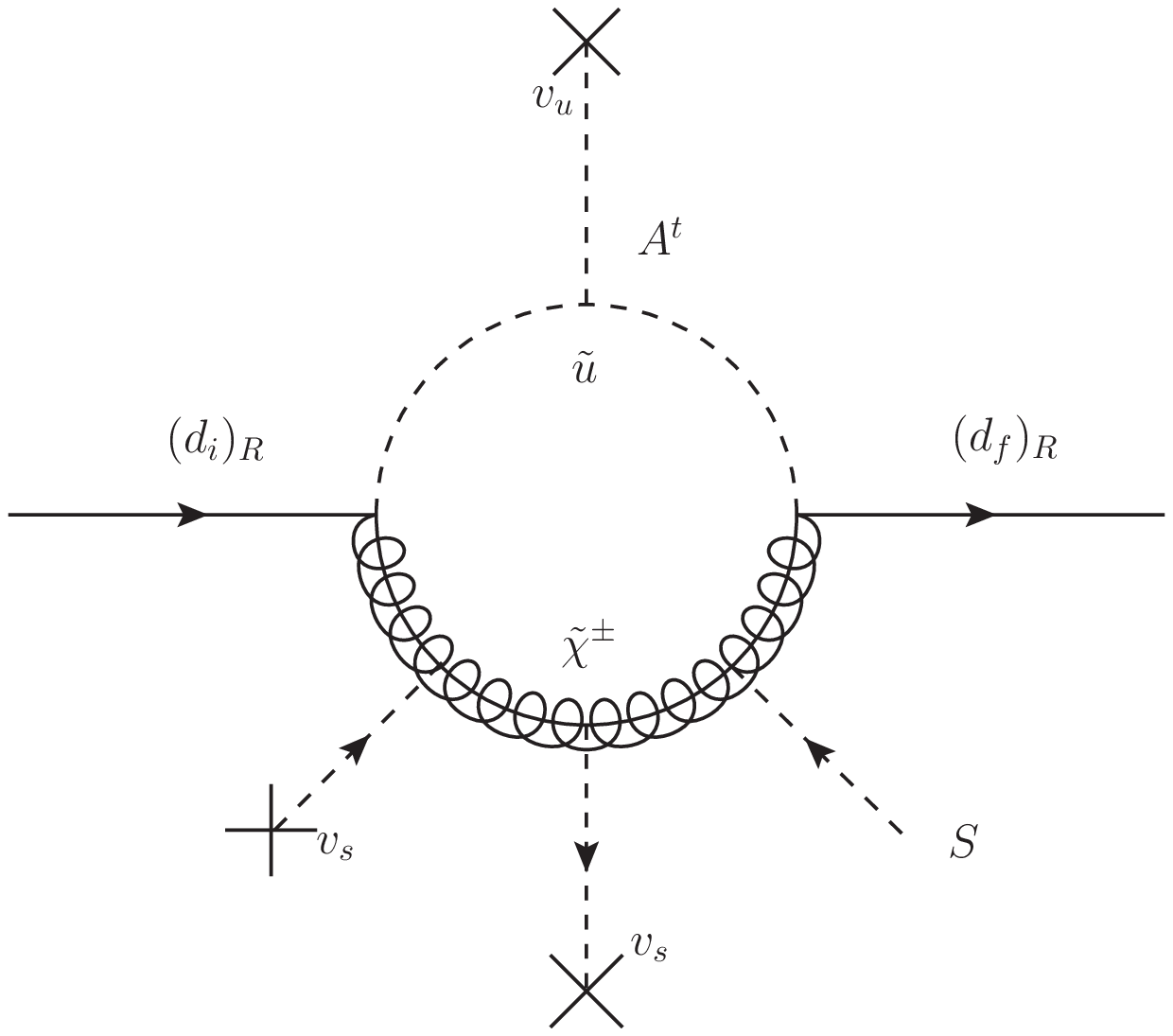}
\caption{Example Feynman diagrams showing the chargino contribution 
to the effective singlet-down-quark couplings in the decoupling limit 
(i.e. in leading order of $v/m_{\rm SUSY}$).}
\label{fig:chargino}
\end{figure}

Concerning the effective couplings induced by loop-diagrams with
 chargino and the neutralinos the situation is more involved since
 $\mu_{\rm eff}$ can also appear from the diagrams like the ones shown
 in Fig.~\ref{fig:chargino}. This effect is contained to all orders
 within the higgsino propagator emerging from the Dyson series
\begin{equation}
\dfrac{1}{{\cancel{k}}} + \dfrac{1}{{\cancel{k}}}\,\mu_{\rm eff} \,
\dfrac{1}{{\cancel{k}}} + \dfrac{1}{{\cancel{k}}}\,\mu_{\rm eff}\, 
\dfrac{1}{{\cancel{k}}}\,\mu _{\rm eff}^*\, \dfrac{1}{{\cancel{k}}} +
 \dfrac{1}{{\cancel{k}}}\,\mu _{\rm eff}\, \dfrac{1}{{\cancel{k}}}\,
\mu _{\rm eff}^*\, \dfrac{1}{{\cancel{k}}}\,\mu _{\rm eff} \,
\dfrac{1}{{\cancel{k}}}+... = 
\dfrac{{\cancel{k} + \mu _{\rm eff} }}{{k^2  - \left| {\mu _{\rm eff}^2 } 
\right|}}\,.
\end{equation}
Each $\mu _{\rm eff}=\lambda v_s$ arises from a coupling to the
 singlet. In the corresponding self-energy in the decoupling
 limit~(\eq{neutralino_SE} and~\eq{chargino-SE}) we pick out only
 the part of the propagator proportional to $\mu_{\rm eff}$~(or
 $\mu_{\rm eff}^*$) meaning that we necessarily have an odd number of
 couplings to the singlet whose contributions to the Dyson series is:
\begin{equation}
\dfrac{1}{{\cancel{k}}}\,\mu _{\rm eff} \,\dfrac{1}{{\cancel{k}}} 
+ \dfrac{1}{{\cancel{k}}}\,\mu _{\rm eff}\, \dfrac{1}{{\cancel{k}}}\,
\mu _{\rm eff}^*
 \,\dfrac{1}{{\cancel{k}}}\,\mu _{\rm eff}\, \dfrac{1}{{\cancel{k}}} + ... = 
\dfrac{{\mu _{\rm eff} }}{{k^2  - \left| {\mu _{\rm eff}^2 } \right|}}\,.
\end{equation}
In all except one of these couplings the singlet is replaced by its 
 VEV $v_s$. If the coupling is to $S$ (and not $S^*$) the Dyson series gives 
\begin{equation}
\dfrac{1}{{\cancel{k}}}\,\lambda S\,\dfrac{1}{{\cancel{k}}} 
+ 2\dfrac{1}{{\cancel{k}}}\,\lambda S\,\dfrac{1}{{\cancel{k}}}\,
\mu_{\rm eff}^*\, \dfrac{1}{{\cancel{k}}}\,\mu_{\rm eff} \,
\dfrac{1}{{\cancel{k}}} 
+ ... = \dfrac{{k^2 \lambda S}}{{\left( {k^2  - \left| {\mu _{\rm eff}^2 } 
\right|} \right)^2 }} = \dfrac{{\lambda S}}{{k^2  - \left| {\mu _{\rm eff}^2 } 
\right|}} + \dfrac{{\left| {\mu_{\rm eff}^2 } \right|\lambda S}}{{\left( {k^2  
- \left| {\mu _{\rm eff}^2 } \right|} \right)^2 }} \, . 
\end{equation}
where the factor 2 takes into account the possible permutations. 
If the coupling is to $S^*$, on the other hand, we have instead 
\begin{equation}
\dfrac{ \mu ^2 \lambda^* S^* } 
 {{\left( {k^2 - \left| {\mu _{\rm eff}^2 } \right|} \right)^2 }} \, . 
\end{equation}
Thus we can write for the chargino induced quark-singlet coupling in the 
following way:
\begin{equation}
\renewcommand{\arraystretch}{2.4}
  \begin{array}{l}
 \widehat\Gamma _{d_f d_i }^{\tilde \chi ^ \pm  h_s \,LR}  = 
-\dfrac{1}{\sqrt{2}}\left( {\lambda \dfrac{{ v_u }}{{\mu _{\rm eff} }} 
E_{fi}^{\prime d\tilde \chi ^ \pm  } 
 + 2\mu _{\rm eff} {\mathop{\rm Re}\nolimits} \left[ {\mu_{\rm eff}^* \lambda } 
\right]
 \dfrac{\partial }{{\partial \left| {\mu _{\rm eff}^2 } \right|}}
 \left( {\dfrac{{ v_u  }}{{\mu _{\rm eff} }}} E_{fi}^{\prime d\tilde \chi ^ \pm }
 \right)} \right)\,,\\
 \widehat\Gamma _{d_f d_i }^{\tilde \chi ^ \pm  a_s LR}  = 
 -\dfrac{1}{\sqrt{2}}\left( {\lambda \dfrac{{ v_u }}{{\mu _{\rm eff} }}
E_{fi}^{\prime d\tilde \chi ^ \pm }  
 + 2i\mu _{\rm eff} {\mathop{\rm Im}\nolimits} 
\left[ {\mu_{\rm eff}^* \lambda }  \right]
 \dfrac{\partial }{{\partial \left| {\mu _{\rm eff}^2 } \right|}}
 \left( {\dfrac{{ v_u }}{{\mu _{\rm eff} }}}  
E_{fi}^{\prime d\tilde \chi ^ \pm }   \right)} \right)\,.
 \end{array}
\label{S_chargino}
\end{equation}
Of course also the lepton-singlet coupling induced by charginos can be written 
in the same way and the corresponding formula for the part of the neutralino 
self-energy which contains gaugino-higgsino mixing is straightforward. 
Note that the formulas (\ref{S_chargino}) are also valid for gluino and 
bino contributions, where the second term vanishes.

We denote the sum of all contributions as:
\begin{equation}
\widehat\Gamma _{d_f d_i }^{a_s LR}  = 
\widehat\Gamma _{d_f d_i }^{\tilde \chi ^ \pm  a_s LR}
+\widehat\Gamma _{d_f d_i }^{\tilde \chi ^ 0  a_s LR}
+\widehat\Gamma _{d_f d_i }^{\tilde g  a_s LR}\,,\qquad
 \widehat\Gamma _{d_f d_i }^{h_s LR}  
= \widehat\Gamma _{d_f d_i }^{\tilde \chi ^ \pm  h_s LR}
+\widehat\Gamma _{d_f d_i }^{\tilde \chi ^ 0  h_s LR}
+\widehat\Gamma _{d_f d_i }^{\tilde g  h_s LR}\,.
\end{equation}
and for leptons $d$ is simply replaced by $\ell$.

\subsection{Higgs couplings in the physical basis}

Until now, we calculated the effective couplings of the Higgs doublets
 and the singlet to matter fermion. For this we worked in the
 interaction eigenstate for the Higgs sector and in the super-CKM basis
 with diagonal Yukawa couplings for the quarks and leptons. 
Both for quarks/leptons and for the Higgses, this is not the physical basis 
with diagonal mass matrices. Therefore, additional rotations 
(see \eq{Higgrotations} and \eq{DeltaU}) are required. 
Note that since the masses of the quarks and leptons entirely originate from the
 doublets $H_u$ and $H_d$ their relation to the Yukawa couplings is the same 
as in the MSSM. 

After integrating out the heavy SUSY particles, interactions of the doublet Higgs bosons with quarks are described by the Lagrangian
\begin{equation}
\renewcommand{\arraystretch}{1.8}
\begin{array}{l}
\mathcal{L}_{\rm doublet}^{\rm sCKM} = \bar{Q}^a_{f\,L} \left[
  \left(Y^{d}_{i}\delta_{fi}+E^{d}_{fi}\right)\epsilon_{ab}H^{b\star}_d\,-\,
E^{\prime d}_{fi} H^{a}_u \right]d_{i\,R}\\
+ \, \bar{Q}^a_{f\,L} \left[ \left(Y^{u}_{i}\delta_{fi}+E^{u}_{fi} \right)
 \epsilon_{ba} H^{b\star}_u \, - \, E^{\prime u}_{fi} H^{a}_d
  \right]u_{i\,R}\,+ \, {h.c.} \; ,
  \end{array}
  \label{Leff}
\end{equation}
with $E^q$ and $E^{\prime q}$ determined by Eqs.~(\ref{HoloDeco}) and~(\ref{HoloDeco-up}).  
After EW symmetry breaking the quark mass matrices are given by:
\begin{eqnarray}
m^d_{fi}&=v_d\left(Y^{d}_{i}\delta_{fi}+E^{d}_{fi}\right)
+v_u E^{\prime d}_{fi}\,,\\
m^u_{fi}&=v_u\left(Y^{u}_{i}\delta_{fi}+E^{u}_{fi}\right)
+v_d E^{\prime u}_{fi}\,.
\end{eqnarray}
We use these relations in order to eliminate the dependence on $Y^{q}_{i}$ and $E^{q}_{fi}$ in \eq{Leff}.
In addition, we go to the physical basis with diagonal quark mass matrices
\begin{equation}
	U^{q\,L\star}_{jf} m^q_{jk}U^{q\,R}_{ki}=m_{q_i}\delta_{fi}\, .
	\label{mphys}
\end{equation}
In the physical basis with diagonal quark mass matrices, the doublet-Higgs interactions with quarks are given by
\begin{eqnarray}
\mathcal{L}_{\rm doublet} = &-& \bar d_{f\,L} \left[\left(\dfrac{{m_{d_i }
  }}{{v_d }}\delta_{fi} - \epsilon_{fi}^{ d}\tan\beta
  \right)H_d^{0\star}\,+\,\epsilon_{fi}^{ d}\,H_u^0 \right]d_{i\,R}
\nn \\
&-& \bar u_{f\,L} \left[\left(\dfrac{{m_{u_i } }}{{v_u }}\delta_{fi} -
  \epsilon_{fi}^{ u}\cot\beta \right)H_u^{0\star}\,
+\,\epsilon_{fi}^{ u}\,H_d^{0} \right] u_{i\,R} \nn\\
&+& \bar u_{f\,L} V_{fj} \left[ {\dfrac{{m_{d_i } }}{{v_d
    }}\delta_{ji}-\left( {\cot \beta + \tan \beta }
    \right) \epsilon_{ji}^{ d}  } \right]H^{+}_d\ d_{i\,R} \nn \\
&+& \bar d_{f\,L} V_{jf}^{\star} \left[ { \dfrac{{m_{u_i
    } }}{{v_u }}\delta_{ji}-\left( {\tan \beta +
      \cot\beta } \right)\epsilon_{ji}^{ u}  } \right] H^{-}_u u_{i\,R}\,
+\,{h.c.}  \,\,\, .
\label{L-Y-FCNC}
\end{eqnarray}
Here we defined the quantity\footnote{We have dropped most of the parts 
contributing to $E^{\prime u}$, and therefore $\epsilon^u$, 
in Eq. (\ref{up-SE}), since their contributions to 
the singlet Higgs couplings are suppressed by $\cot\beta$ and 
irrelevant for our study. See, for example, 
Refs.~\cite{Degrassi:2000qf,Carena:2000uj,Crivellin:2011jt} 
for the contributions of $\epsilon^u$ to the effective couplings of $H^{\pm}$ 
to up-type quarks.} 
\begin{equation}
\epsilon_{fi}^{f} \equiv (U^{fL\dagger} E^{\prime f} U^{fR} )_{fi} . 
\end{equation}
which gives rise to the chirally-enhanced corrections to the Higgs couplings 
in the physical basis. In the Lagrangian
\begin{eqnarray}
L_{\rm eff} &=& \Gamma_{q_fq_i}^{H^0_k\,LR} H^0_k \bar{q}_{fL} q_{iR}
+i \Gamma_{q_fq_i}^{A^0_k\,LR} a^0_k \bar{q}_{fL} q_{iR}
\nonumber \\
&& 
+\Gamma_{u_fd_i}^{H^\pm\,LR} H^+ \bar{u}_{fL} d_{iR}
+\Gamma_{d_fu_i}^{H^\pm\,LR} H^- \bar{d}_{fL} u_{iR}\,+\,(h.c.) \, ,
\end{eqnarray}
with $H_k^0=(h_d, h_u)$ and $A_k^0=(A^0, G^0)$, 
this leads to the following effective Higgs couplings to down-type quarks
\begin{eqnarray}
{\Gamma_{d_f d_i }^{h_d\,LR} } &=& -\dfrac{1}{\sqrt{2}} 
\left( \frac{m_{d_i}}{v_d} \delta_{fi} - \epsilon_{fi}^{d}\tan\beta \right) 
\,,
\nonumber \\[0.1cm]
{\Gamma_{d_f d_i }^{h_u \,LR} } &=& -\dfrac{1}{\sqrt{2}}
\epsilon_{fi}^{ d} \,,\nonumber \\[0.1cm]
{\Gamma_{d_f d_i }^{A^0 \,LR} } &=& \dfrac{1}{\sqrt{2}}\sin\beta 
\left( \frac{m_{d_i
}}{v_d} \delta_{fi} - \epsilon_{fi}^{d}\tan\beta \right) \,, 
\label{Higgs-vertices-decoupling}
\end{eqnarray}
while the couplings of the Nambu-Goldstone boson $G^0$ 
${\Gamma_{d_f d_i }^{G^0 \,LR} }=(1/\sqrt{2})\cos\beta (m_{d_i}/v_d) 
\delta_{fi}$ receive no corrections from $\epsilon^d$. 

For leptons the neutral Higgs vertices follow trivially from the ones
for down-type quarks.

The quark field rotations in \eq{mphys} also lead to a redefinition of 
the quark-singlet couplings from the super-CKM basis as 
\begin{align}
\Gamma _{d_f d_i }^{h_s(a_s) \,LR}&=U^{dL*}_{f^\prime f}
\widehat\Gamma _{d_{f'} d_{i'} }^{h_s(a_s) \,LR}U^{dR}_{i^\prime i}\,,\\
\Gamma _{\ell_f \ell_i }^{h_s(a_s) \,LR}&=U^{\ell L*}_{f^\prime f}
\widehat\Gamma _{\ell_{f'} \ell_{i'} }^{h_s(a_s) \,LR}
U^{\ell R}_{i^\prime i}\,.
\end{align}

Finally, we obtain the following couplings to the Higgs mass eigenstates:
\begin{eqnarray}
\left( \begin{array}{c} \Gamma _{d_f d_i }^{h_2\,LR} \\ 
\Gamma _{d_f d_i }^{h_3\,LR} \\ \Gamma _{d_f d_i }^{h_1\,LR} 
\end{array} \right) 
= \left( O^S \right)^T\left( \begin{array}{c} \Gamma _{d_f d_i }^{h_d
\,LR} \\ \Gamma _{d_f d_i }^{h_u\,LR} \\ \Gamma _{d_f d_i }^{h_s\,LR} 
\end{array} \right)
\, ,\qquad
 \left( \begin{array}{c} \Gamma _{d_f d_i }^{G^0\,LR} \\ 
\Gamma _{d_f d_i }^{a_2\,LR} \\ \Gamma _{d_f d_i }^{a_1\,LR} \end{array} 
\right) = \left(  O^P_{\theta_A} \right)^T\left( 
\begin{array}{c} \Gamma _{d_f d_i }^{G^0\,LR} \\ 
\Gamma _{d_f d_i }^{A^0\,LR} \\ 
\Gamma _{d_f d_i }^{a_s\,LR} \end{array} \right) \, .
\label{coupling_in_mass}
\end{eqnarray}
The analogous results for charged leptons follow by replacing $d$ with $\ell$.

At this point we can already make a rough estimation of 
the effective singlet couplings to down-type quarks 
$\Gamma_{d_f d_i}^{(h_s,a_s)LR}$ (and charged leptons as well). 
Comparing Eq. (\ref{S_chargino}) and Eq. (\ref{Higgs-vertices-decoupling}), 
it is seen that the singlet couplings are suppressed by $v/v_s$ compared to the loop-induced couplings of 
the doublets ($h_d$, $A^0$). As a result, the singlet couplings decouple with 
$(M_{\rm SUSY},\mu_{\rm eff})\to \infty$ and fixed $\lambda$, while 
the effective doublet Higgs boson couplings remain finite in the same limit. 
Furthermore, the scaling with $\tan\beta$ is different: 
while the loop-induced parts of the doublet couplings scale 
as $\tan^2\beta$, the singlet couplings scale as $\tan\beta$. 
Nevertheless, for the Higgs mass eigenstates which are almost pure singlet, as is the case for example for pseudo-axions in 
Peccei-Quinn symmetric limit (\ref{eq:cosTHA}), 
$\Gamma^{h_s,a_s}$ may give dominant contribution to the 
effective couplings to down-type quarks. 

\section{Numerical Results}

In this section we study the numerical significance of loop-induced effective 
couplings of the singlet Higgs to down-type quarks and charged leptons. 
We first illustrate the generic size of these couplings and then 
discuss the potential effect of the singlet-induced contributions to 
flavour physics. 

In the following we will quantify the size of the effective couplings 
of singlet Higgs bosons to down-type quarks (charged leptons) 
$\Gamma^{(h_s,a_s)\;LR}_{d_f d_i}$ ($\Gamma^{(h_s,a_s)\;LR}_{\ell_f \ell_i}$). 
For the couplings to quarks we include QCD running and 
evaluate them at the squark mass scale. 
As input values for SM parameters ($\alpha_s$, $m_b$, \ldots) we use the 
 current PDG values~\cite{PDG}.
For the SUSY sector parameters, we use the following values: $M_1=M_2=1$ TeV, 
$m_{\tilde{g}}=3$ TeV, $\lambda=1$. 
The chirality-conserving submatrix of the sfermion mass matrix 
$\mathcal{M}^2_f$ in Eq.~(\ref{eq:Deltas}) is assumed to be 
diagonal and flavour-independent, $m_{\tilde{q}}=2$ TeV and $m_{\tilde{\ell}}=1$ TeV 
for squarks and sleptons, respectively. 
In the discussion of the flavour-changing couplings, 
however, we introduce the mixing  between the third and second generations 
in the left-left or right-right sector. 
The chirality-conserving submatrix is then expressed as 
\begin{equation}
m^2_{dLL} = m_{\tilde{q}}^2 \left( \begin{array}{ccc} 1 & 0 & 0 \\
0 & 1 & \delta^{dLL}_{23} \\ 0 & \delta^{dLL}_{23} & 1 
\end{array} \right) \,  ,
\end{equation}
while $m^2_{\ell LL}$ and $m^2_{fRR}$ ($f=u,d,\ell$) are obtained 
by replacing $\delta^{dLL}_{23}$ by  $\delta^{\ell LL}_{23}$ and 
$\delta^{fRR}_{23}$, respectively (and $m_{\tilde{q}}\to m_{\tilde{\ell}}$ for sleptons).  
The A-parameters in the chirality-changing submatrices are set to 
\begin{equation}
(A^u)_{ij} = {\rm diag}(0,0,A^t), \; 
(A^d)_{ij} = {\rm diag}(0,0,A^b), \; 
(A^{\ell})_{ij} = {\rm diag}(0,0,A^{\tau}).  
\end{equation}
We use $A^b=A^{\tau}=1$ TeV. 
Other parameters are specified in the figure captions. 
As stated before, we consider the case where CP 
violation is negligible and set all parameters to real. 

\subsection{Flavour-conserving couplings}

We first show the flavour-conserving effective couplings of the 
singlet Higgs bosons to the down-type quarks and charged leptons, 
for the case of minimal flavour violation (MFV) 
($\delta_{23}^{fLL}=\delta_{23}^{fRR}=0$). 
Since these couplings are purely loop-induced, 
they are much smaller than 
the couplings to the doublet Higgs, which exists already at the tree-level. 
This is especially the case for charged leptons where the loop corrections 
cannot involve the strong interaction. 

As shown in Eq.~(\ref{S_chargino}), there are two types of the 
contributions to the singlet couplings: 
the one directly proportional to the non-holomorphic part of the self energy 
$E^{\prime f \,LR}$ and the other involving the derivative 
with respect to $\mu_{\rm eff}$. 
For CP-conserving case, the latter only contribute to the couplings 
of the CP-even component $h_s$, causing the different behavior 
between $\Gamma^{h_s LR}$ and $\Gamma^{a_s LR}$. 

\begin{figure}
\includegraphics[width=.49\linewidth]{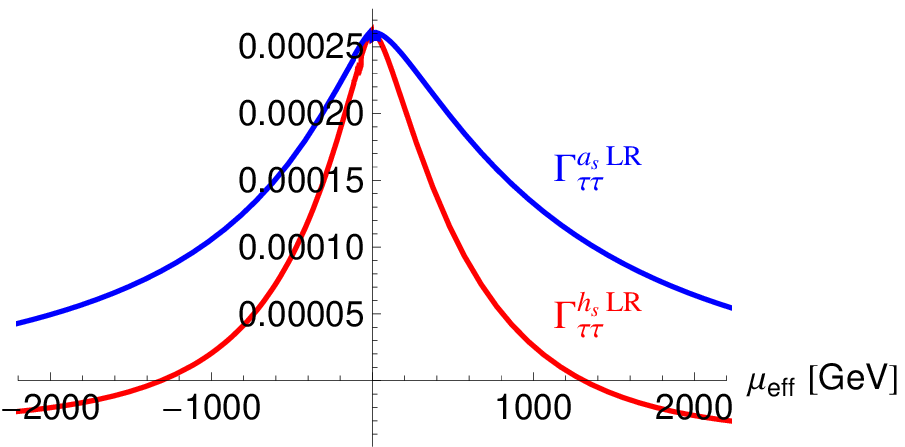}
\includegraphics[width=.49\linewidth]{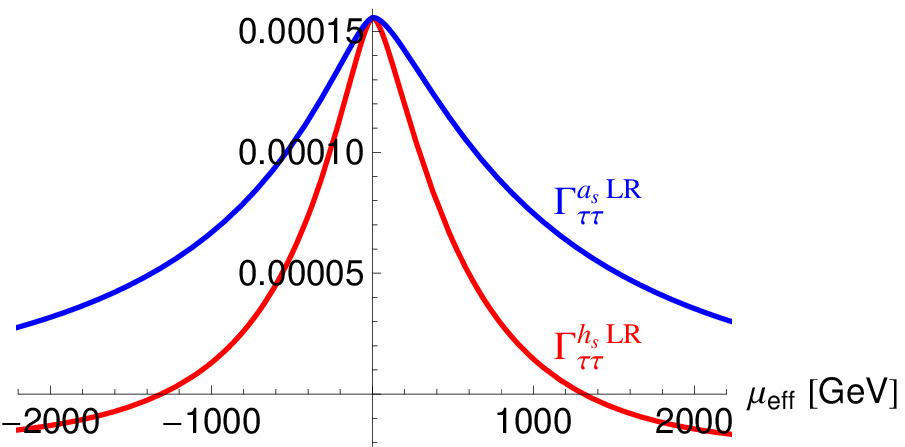}
\caption{Left: Strength of the loop-induced coupling of tau leptons 
to the singlet Higgs for 
$\tan\beta=50$ and $\lambda=1$ as a function of the 
(effective) $\mu$ parameter. Blue (dark gray): $h_s$, red (light gray): 
$a_s$. Right: Same as the left plot for $\tan\beta=30$.}
\label{Fig:Gamma_l}
\end{figure}

Fig.~\ref{Fig:Gamma_l} shows the loop-induced singlet couplings to 
tau leptons. As expected, these couplings are much smaller than the 
Higgs-tau coupling in SM, $m_{\tau}/(\sqrt{2}v)=0.0072$. 

\begin{figure}
\includegraphics[width=.49\linewidth]{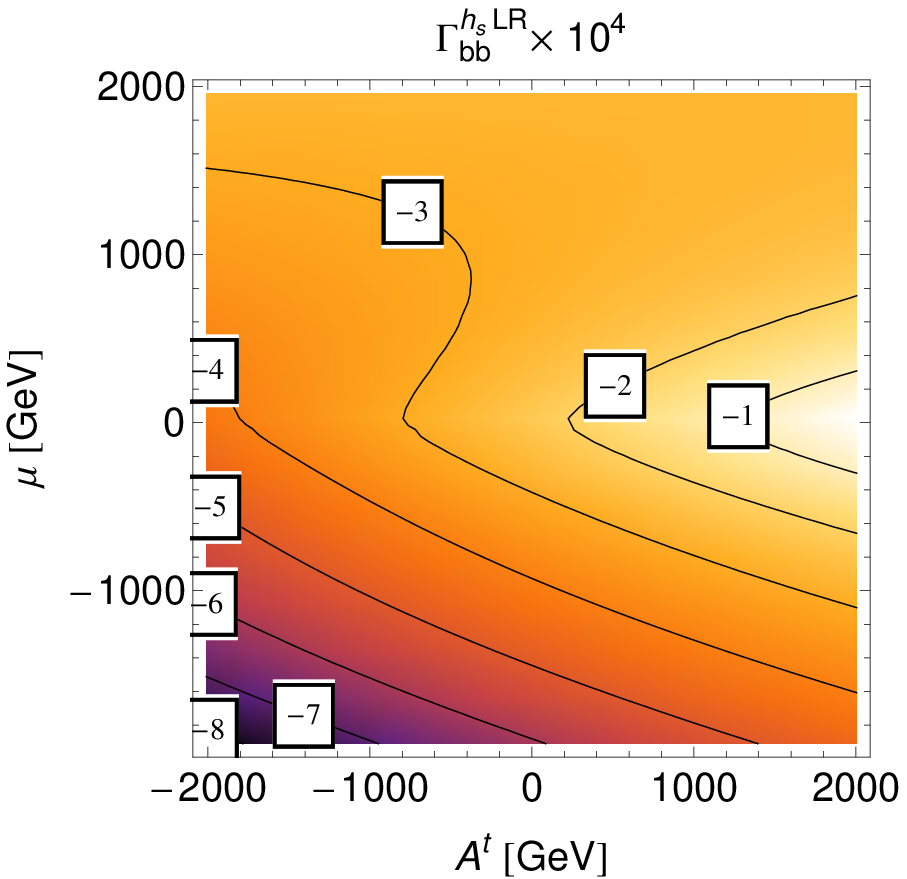}
\includegraphics[width=.49\linewidth]{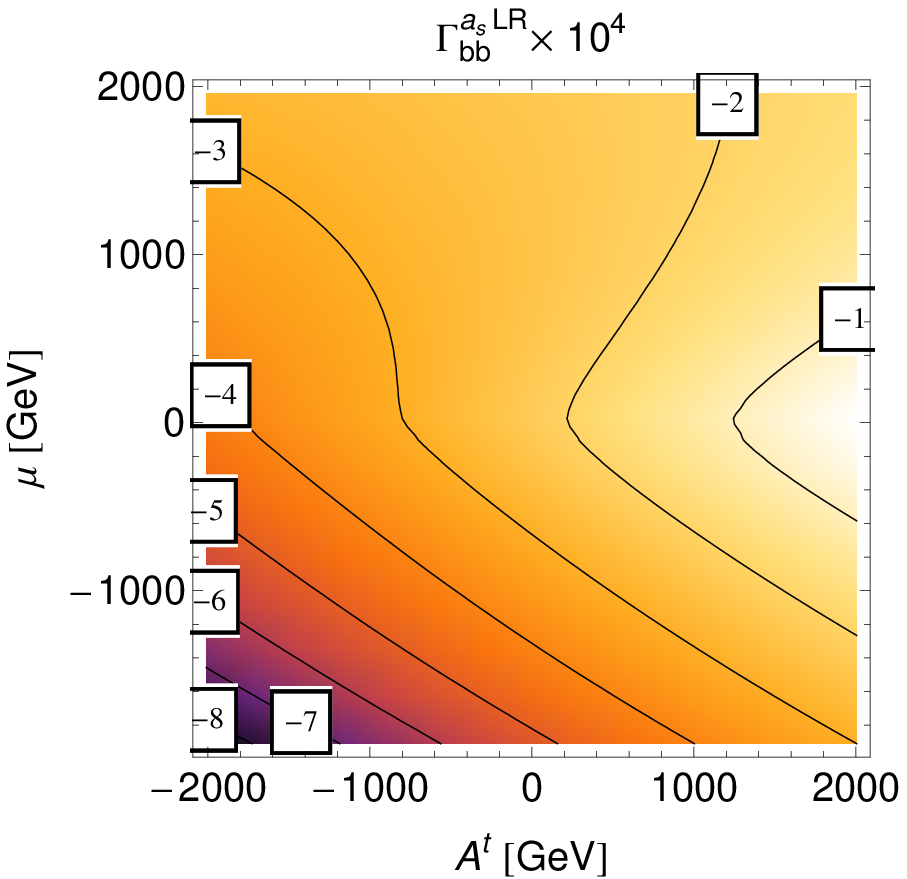}
\caption{Left: Strength of the loop-induced coupling of bottom quarks to the 
singlet Higgs bosons, to $h_s$ (left plot) and to $a_s$ (right plot), for 
$\tan\beta=50$ and $\lambda=1$ in the $\mu-A^t$ plane.}
\label{Fig:at-mu}
\end{figure}

For down-type quarks, the behavior of the effective couplings 
is similar to that for leptons, but two additional numerically 
important parameters enter (beside that evidently squark masses and slepton 
masses enter): $A^t$ and $m_{\tilde g}$. Furthermore, 
the threshold correction to the relation between $Y^{b(0)}$ and $m_b$ is 
much larger than that for $\tau$ lepton (because $\alpha_s$ is involved) 
which leads to an asymmetric behavior of 
$\Gamma^{(h_s,a_s)\,LR}_{bb}$ with respect to the sign of $\mu_{\rm{eff}}$. 
We show the dependence of the effective singlet coupling to 
bottom quarks $\Gamma^{(h_s,a_s)\,LR}_{bb}$ on the 
($A^t$, $\mu_{\rm eff}$) plane in Fig.~\ref{Fig:at-mu}. 
Again, they are smaller than the coupling of the SM Higgs, 
$m_b/(\sqrt{2}v)$(at 2 TeV)$\sim 0.0095$. 

\subsection{Flavour-changing couplings}

The flavour-changing couplings of fermions to the singlet Higgs and to the 
doublet Higgs are both loop-induced. 
As shown in Section~5,
the singlet couplings are suppressed by the factor 
$\sim (v/v_s)\cot\beta$ compare with the corresponding doublet couplings. 
Here we focus on the flavour-changing couplings between the second and
third generations, induced either by the flavour mixing of left-handed
sfermions $\delta^{f\,LL}_{23}$ or by the CKM matrices in
chargino-sfermion loops.

\subsubsection{Leptons}

\begin{figure}
\includegraphics[width=.49\linewidth]{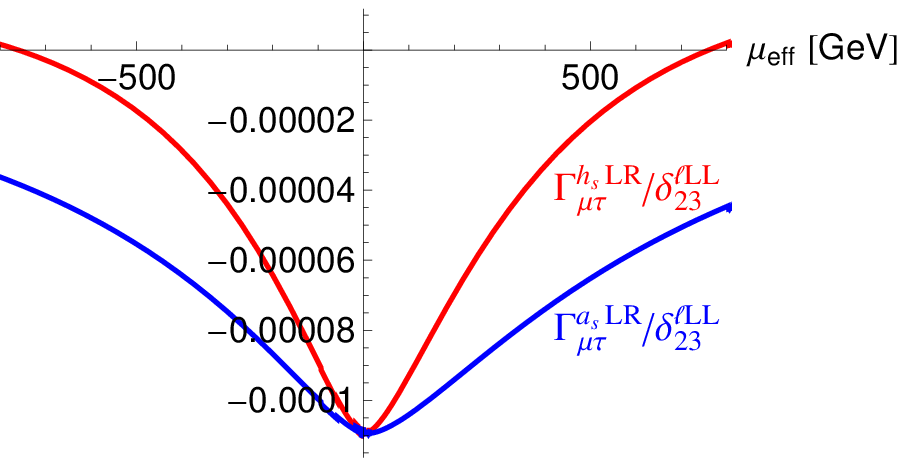}
\includegraphics[width=.49\linewidth]{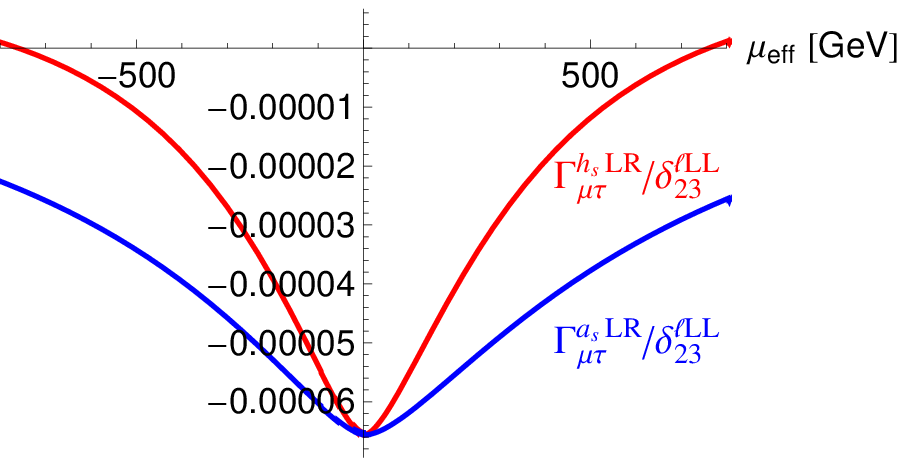}
\caption{Left: Strength of the loop-induced coupling of tau and muon to 
the singlet Higgs 
induced by $\delta^{\ell\;LL}_{23}$ (and normalized to 
$\delta^{\ell\;LL}_{23}$) 
for $\tan\beta=50$ and $\lambda=1$ as a function of the (effective) $\mu$ 
parameter. Blue (dark gray): $h_s$, red (light gray): $a_s$.\newline
Right: Same as left plot for $\tan\beta=30$.}
\label{Fig:Gamma_l23LL}
\end{figure}

The plots of Fig.~\ref{Fig:Gamma_l23LL} show the behavior of effective 
coupling of the singlet Higgs $\Gamma^{(h_s,a_s)\,LR}_{\mu\tau}$ induced by 
$\delta^{\ell\,LL}_{23}$. The contribution is to a very good approximation 
proportional to $\delta^{\ell\;LL}_{23}$, unless $\delta^{\ell\,LL}_{23}$ 
is very large. 
The effect of $\delta^{\ell\;RR}_{23}$ to $\Gamma^{(h_s,a_s)\,LR}_{\mu\tau}$ 
is always suppressed by the ratio $m_{\mu}/m_{\tau}$ 
compared to the $\delta^{\ell\;LL}_{23}$ contribution, 
and therefore in most scenarios subleading. This is also the case 
for the contribution of $\delta^{\ell\;RR}_{23}$ 
to $\Gamma^{(h_s,a_s)\,LR}_{\tau\mu}$ since it does not involve 
the $\widetilde{W}$ loops. 

\subsubsection{Quarks}

\begin{figure}
\includegraphics[width=.49\linewidth]{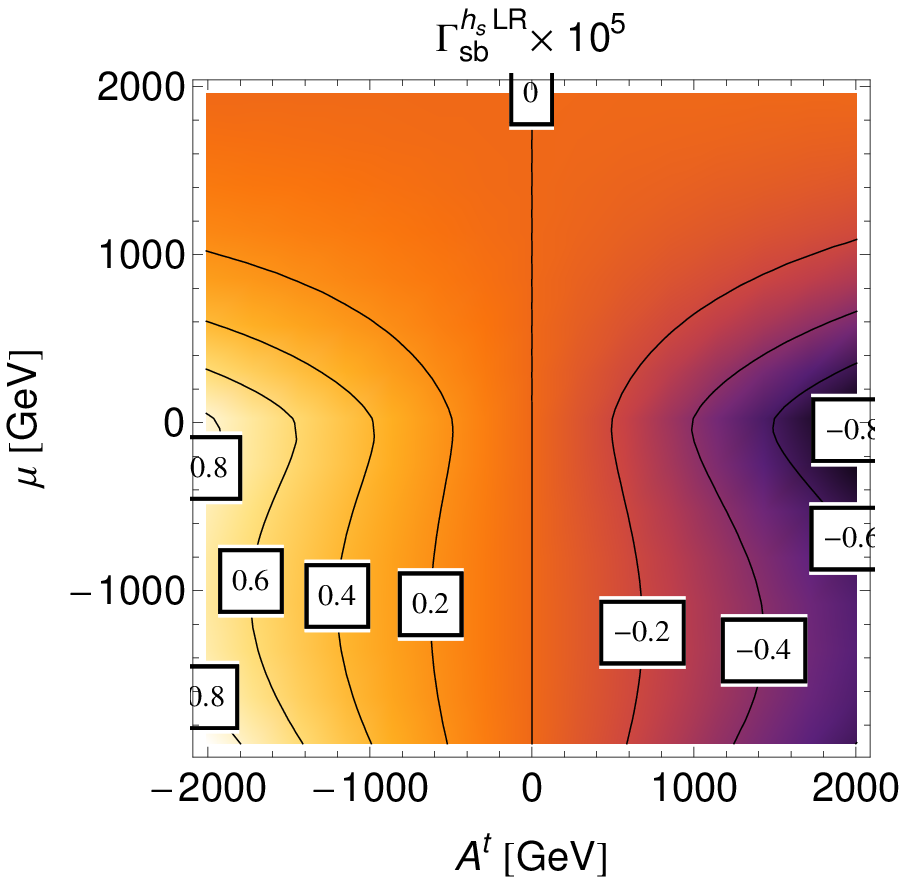}
\includegraphics[width=.49\linewidth]{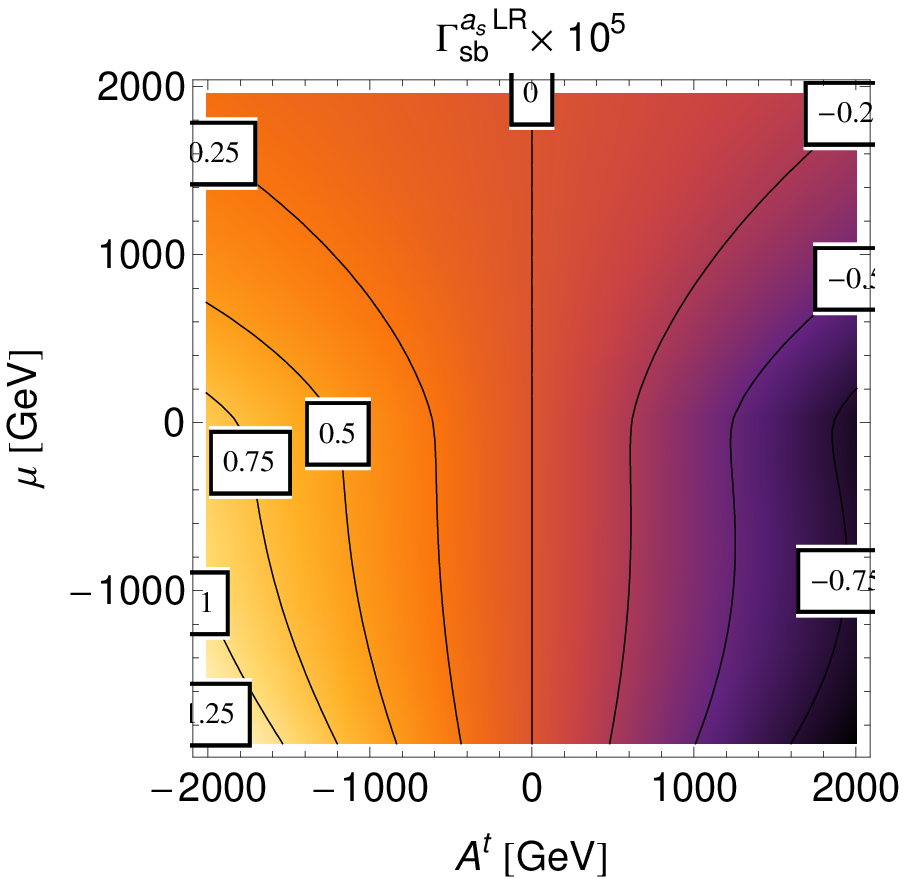}
\caption{Size of the effective singlet couplings to bottom and strange 
quarks for MFV in the $\mu_{\rm eff}$-$A^t$ plane, for 
$\tan\beta=50$ and $\lambda=1$.}
\label{Fig:MFV}
\end{figure}

Let us consider first the case of the MFV ($\delta^q_{23}=0$). 
In this case only the quark-squark-chargino vertex induces flavour-violation. 
In Fig.~\ref{Fig:MFV} we show the size of the effective singlet 
couplings  $\Gamma^{(h_s,a_s)\,LR}_{sb}$ induced via chargino loops. 

\begin{figure}
\includegraphics[width=.49\linewidth]{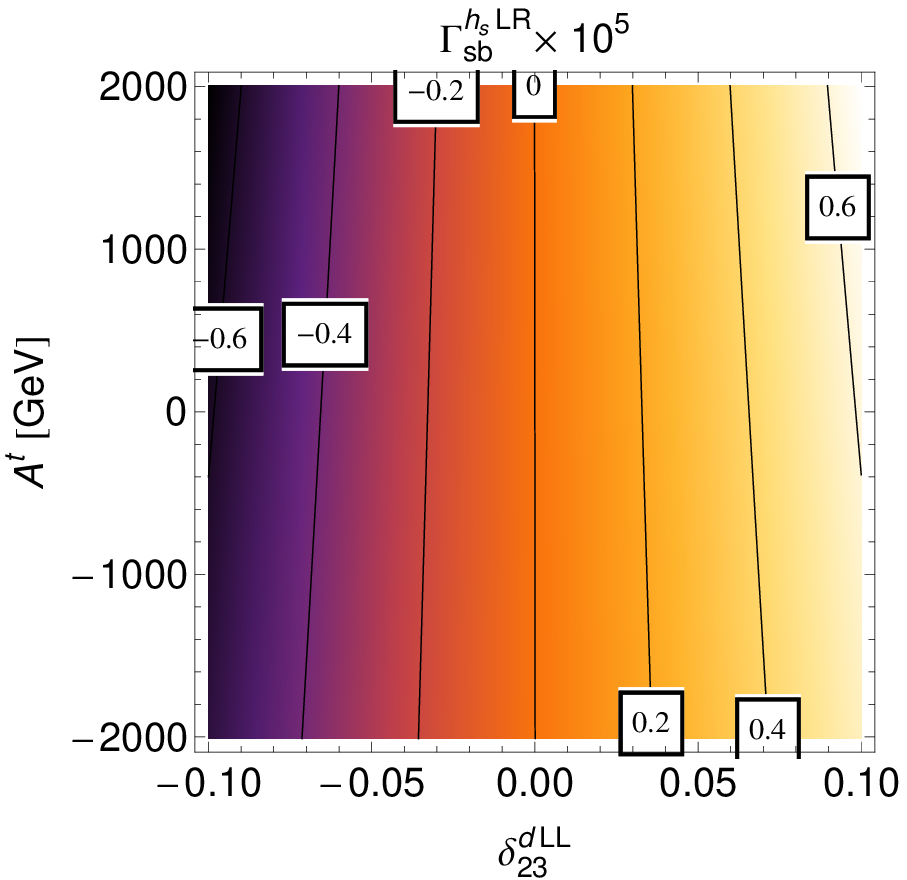}
\includegraphics[width=.49\linewidth]{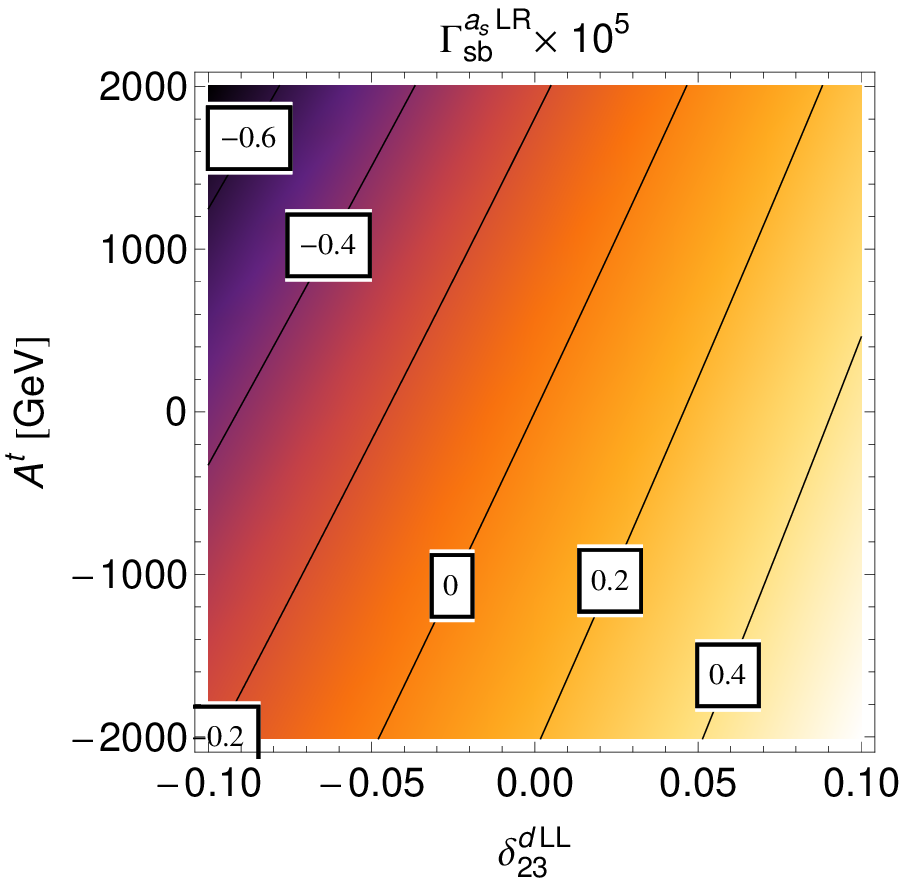}
\caption{Size of the effective singlet couplings to bottom and strange 
quarks 
in the $A^t$-$\delta^{d\,LL}_{23}$ plane for 
$\mu=2$ TeV, $\tan\beta=50$ and $\lambda=1$.}
\label{deltadLL23}
\end{figure}

In the presence of non-minimal sources of flavour-violation 
in the squark sector, additional contributions are induced. 
In case of a non-vanishing element 
$\delta^{d\,RR}_{23}$, $\Gamma^{(h_s,a_s)\,LR}_{bs}$ is generated.  
Since in this case no interference with the MFV contribution occurs, 
their sizes are proportional to $\delta^{d\,RR}_{23}$. 
In the presence of $\delta^{d\,LL}_{23}$, in contrast, there is interference 
with the MFV contributions for  $\Gamma^{(h_s,a_s)\,LR}_{sb}$, 
as seen in Fig.~\ref{deltadLL23}. 

\subsection{Effects in flavour-changing processes}

In general, the flavour-changing neutral current processes 
which are most sensitive to the exchange of 
neutral scalars which couple to quarks proportionally to their mass 
are $B_{s,d}\to \mu^+\mu^-$ and $B_{s,d}-\bar{B}_{s,d}$ mixing. 
For large $\tan\beta$, these processes receive potentially 
large contributions from the double-penguin diagrams mediated by neutral 
Higgs bosons~\cite{Babu:1999hn,BurasCRS,Domingo:2007dx}. 
However, as discussed 
previously, the effective (flavour-conserving) couplings of the singlet 
Higgs scalars to charged leptons are much smaller than the 
ones of the doublet Higgs scalars. 
This means that the possible effect of the singlet exchange in 
$B_{s,d}\to \mu^+\mu^-$ must be much smaller than the effect of 
the doublet Higgs exchange. 
For $B_{s,d}-\bar{B}_{s,d}$ mixing, in contrast, 
two loop-induced flavour-changing Higgs-quark couplings are involved 
in both the doublet and singlet exchanges. As a result, 
singlet contributions scales compared to the doublet contribution as 
$v^2m_A^2/(v_s^2m_{a_s}^2\tan^2\beta)$ ($m_A$ is the typical 
mass of the heavier doublet Higgs bosons), 
which can be non-negligible for very light $a_s$. 
Furthermore, in the MFV case 
the doublet Higgs contribution to $B_{s,d}-\bar{B}_{s,d}$ mixing 
tends to chancel between the CP-even and CP-odd states 
having similar masses~\cite{Babu:1999hn,BurasCRS}, while 
such a suppression is not efficient for the singlet states 
where $h_s$ and $a_s$ can have very different masses.

\begin{figure}
\includegraphics[width=.49\linewidth]{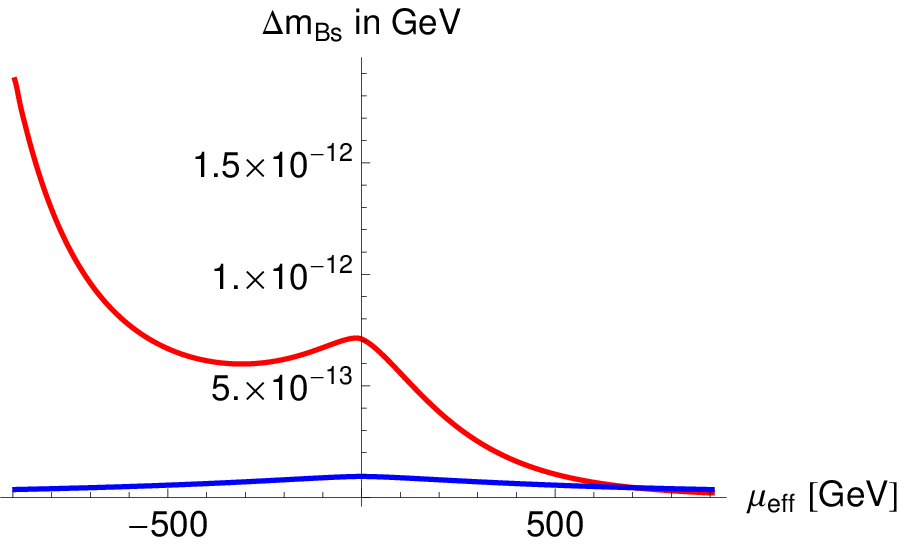}
\includegraphics[width=.49\linewidth]{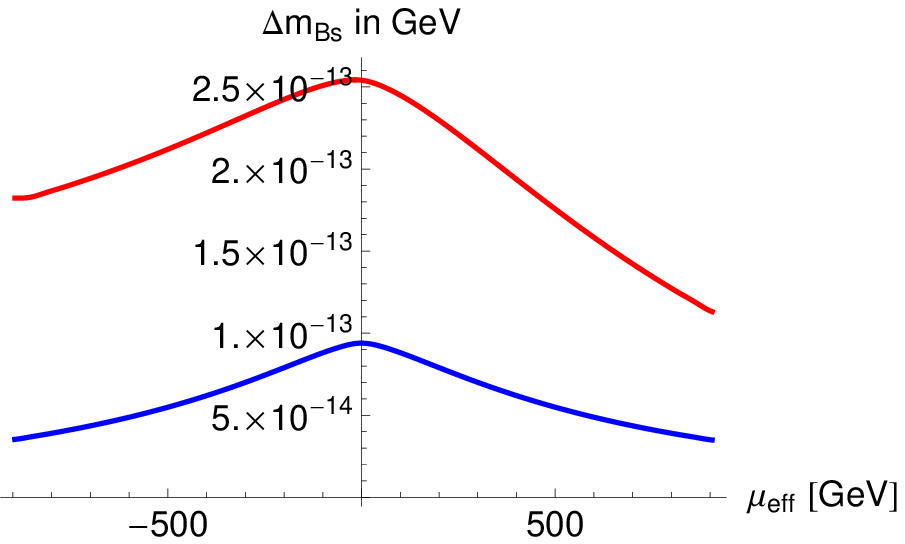}
\caption{Contributions to the mass difference 
in the $B_s$ system generated by chargino 
box diagrams (blue) and by the singlet exchange (red) for 
$\tan\beta=50$ (left plot) and $\tan\beta=30$ (right plot) for 
$\lambda=1$. We assume the singlet-like $a_s$ 
has a mass of 7 GeV while $h_s$ and doublet Higgs bosons 
are much heavier. We set $A^t=3$ GeV. }
\label{BsMixingMFV}
\end{figure}

We illustrate the relative importance of the singlet exchange 
in Fig.~\ref{BsMixingMFV} for the MFV case and $m_{a_s}=7$ GeV. 
The singlet exchange contribution is evaluated by 
using the effective quark-singlet couplings at the 
renormalization scale $m_{a_s}$. The meson form factors 
are adopted from Flavour Lattice Averaging Group 
(FLAG) \cite{Aoki:2013ldr}. 
We see that the singlet Higgs contribution can be much larger than the 
chargino-squark box contribution. 

\section{Conclusion}

In the framework of the MSSM extended by a gauge-singlet supermultiplet 
(e.g the NMSSM), we have studied the loop-induced effective couplings 
of the singlet Higgs bosons to quarks and charged leptons. 
Allowing for the most general flavour structure of the SUSY 
breaking terms, we have derived analytic formula for the couplings of the singlet
to quarks and leptons. Keeping the leading term in the expansion of 
$v/M_{\rm SUSY}$ we have performed the complete resummation of 
all chirally-enhanced effects, making our results valid in the 
phenomenologically large $\tan\beta$ region.

We found that loop-induced singlet couplings to down-type quarks and 
charged leptons are enhanced by one power of $\tan\beta$, while the flavour-changing couplings of down-quarks 
to heavy doublet Higgs bosons are enhanced by $\tan^2\beta$. 
In addition, the loop-induced singlet-quark couplings vanish in the decoupling 
limit $v\ll (M_{\rm SUSY},\mu_{\rm eff})$ 
while the doublet couplings remain finite. 
Nevertheless, the loop-induced singlet couplings can be the dominant 
part of the couplings of the lightest states $a_1$ and/or $h_1$ to quarks and 
charged leptons if they are to a good approximation singlet like. 
Furthermore, these couplings can also be phenomenologically important 
if such $a_1$ is very light as it give enhanced contributions to flavour observables.

While an analysis of the impact of the effective couplings of the singlet 
Higgs in realistic scenarios,  for example in the NMSSM, is beyond 
the scope of this article, we have pointed out that sizable 
effects of singlet Higgs in flavour observables are still possible. 
In our numerical analysis we have examined the generic size of 
the effective singlet quark (lepton) couplings, 
both flavour-conserving and flavour-changing ones, 
and considered their impact on processes where the 
singlet contribution is particularly relevant: 
for very low singlet masses, we have shown that it can be the dominant beyond-SM contribution to the $B_s-\bar{B}_s$ mixing due to the enhancement by 
light mass of $a_s$.

\section*{Acknowledgements}
We are grateful to Francesca Borzumati for initializing this project and 
for collaboration at early stage. We thank Kwang Sik Jeong for useful discussions, and 
Florian Staub, Margarete M\"uhlleitner and Ulrich Ellwanger 
for informing the present status of the codes \cite{NMSSMCALC,SPheno,CODESa}. 
The work of A.~C. was supported by a Marie Curie Intra-European Fellowship of the European Community's 7th Framework Programme under 
contract number PIEF-GA-2012-326948 and by the Swiss National Science Foundation.

\end{document}